\documentclass[twocolumn,aps,prb,superscriptaddress,showpacs,
eqsecnum
]{revtex4}
\usepackage{bm,amsmath,amsfonts,amssymb,esint,graphicx,color}

\newcommand{\Eq}[1]{Eq.~\eqref{#1}}
\newcommand{\eq}[1]{\eqref{#1}}
\newcommand{\Fig}[1]{Fig.~\ref{#1}}
\newcommand{\Sec}[1]{Sec.~\ref{#1}}

\newcommand{\beq}{\begin{equation}}
\newcommand{\eeq}{\end{equation}}
\newcommand{\beqa}{\begin{eqnarray}}
\newcommand{\eeqa}{\end{eqnarray}}
\newcommand{\Beqa}{\begin{eqnarray*}}
\newcommand{\Eeqa}{\end{eqnarray*}}
\newcommand{\nn}{\nonumber}

\newcommand{\pdag}{{\phantom{\dagger}}}

\newcommand{\past}{{\phantom{\ast}}}
\newcommand{\sgn}{\text{sgn}}
\newcommand{\arccosh}{\text{arccosh}}
\newcommand{\negquad}{\mkern-18mu}

\newcommand{\triplecolon}{\raisebox{-0.075ex}{\scalebox{0.7}{\bm\vdots}}}

\def\XXint#1#2#3{{\setbox0=\hbox{$#1{#2#3}{\int}$}
     \vcenter{\hbox{$#2#3$}}\kern-.5\wd0}}

\begin{document}

\title{Fate of classical solitons in one-dimensional quantum systems}

\author{M. Pustilnik}
\affiliation{School of Physics, Georgia Institute of Technology, Atlanta, Georgia 30332, USA}
\author{K. A. Matveev}
\affiliation{Materials Science Division, Argonne National Laboratory, Argonne, Illinois 60439, USA}

\begin{abstract}
We study one-dimensional quantum systems near the classical limit described by the Korteweg-de Vries (KdV) equation. The excitations near this limit are the well-known solitons and phonons. The classical description breaks down at long wavelengths, where quantum effects become dominant. Focusing on the spectra of the elementary excitations, we describe analytically the entire classical-to-quantum crossover. We show that the ultimate quantum fate of the classical KdV excitations is to become fermionic particles and holes. We discuss in detail two exactly solvable models exhibiting such crossover, the Lieb-Liniger model of bosons with weak contact repulsion and the quantum Toda model. We argue that the results obtained for these models are universally applicable to all quantum one-dimensional systems with a well-defined classical limit described by the KdV equation.
\end{abstract}



\maketitle

\section{Introduction}
\label{intro}

The Korteweg-de Vries (KdV) equation\cite{KdV_original} describes propagation of waves in a medium with competing dispersion and nonlinearity and is ubiquitous in the physics of classical nonlinear systems.\cite{Soliton_books} In addition to the usual periodic waves, the KdV equation supports solitons, i.e., localized disturbances with particle-like properties. Protected by the integrability of the KdV equation, the solitons move and scatter off each other without distortion. 

Real systems are never integrable. Nevertheless, the KdV equation often provides an adequate effective description of the excitations in the long-wavelength limit, whereas deviations from the integrability have a significant impact on the lifetime and stability of solitons at shorter wavelengths.\cite{PerturbationTheory} Another limitation on the applicability of the KdV equation, which is the primary focus of this paper, arises in one-dimensional quantum systems with a well-defined classical limit. In these systems quantum effects inevitably become dominant at sufficiently long wavelengths, leading to the breakdown\cite{KKG,LMP,AP,ISG_RMP,PM_LiebLiniger,PM_WignerToda,Protopopov 2014} of the classical description. 

Because the emergence of quantum behavior is not associated with broken integrability, it is natural to approach the problem of describing the classical-to-quantum crossover from the perspective of exactly solvable models.\cite{KBI,Sutherland_book} Fortunately, solvable models exhibiting such crossover are available. In this paper, we consider two well-known examples. The first one is the Lieb-Liniger model of bosons with weak contact repulsion.\cite{Lieb,KBI,Sutherland_book,Yang} In the classical limit, its dynamics can be described by the mean-field Gross-Pitaevskii equation,\cite{GP,Pitaevskii} which has the form of the nonlinear Schr{\"o}dinger equation.\cite{Soliton_books} The equation is integrable\cite{Soliton_books} and reduces to the KdV equation in the long-wavelength limit.\cite{Tsuzuki} The second model we consider is the quantum Toda model.\cite{Sutherland_book,Sutherland,Gutzwiller,Sklyanin} Its classical counterpart, the Toda model,\cite{Toda} is also integrable and the corresponding equation of motion in the continuum limit coincides\cite{Toda} with the KdV equation. 

It is well known that at high momenta the exact spectra of elementary excitations of the Lieb-Liniger and the quantum Toda models match those deduced from their classical analogs.\cite{KMF,Sutherland,Sutherland_book} In this paper we focus on low momenta, and evaluate the spectra near the classical-to-quantum crossover. Some of the results presented below have been reported in Refs.~[\onlinecite{PM_LiebLiniger,PM_WignerToda}].

The rest of the paper is organized as follows: In \Sec{KdV} we summarize some well-known facts about the KdV equation and discuss its applicability for the description of quantum interacting systems near the classical limit. In \Sec{models} we introduce the Lieb-Liniger and the quantum Toda models and discuss their relation to the KdV equation. Spectra of the elementary excitations are evaluated in \Sec{Excitations}. The results are discussed in \Sec{Discussion}. Technical details are relegated to the Appendixes.

\section{KdV equation in quantum systems}
\label{KdV}

The KdV equation in the standard form reads\cite{Soliton_books}
\beq
\partial_\tau f + 6f\partial_\xi f + \partial^3_\xi f = 0,
\label{KdV 1}
\eeq
where $\tau$ and $\xi$ are the dimensionless time and coordinate, respectively. The equation can be viewed\cite{Soliton_books} as a canonical equation of motion $\partial_\tau f = \{f,\mathcal H\}$ generated by the Hamiltonian
\beq
\mathcal H =  \int\!d\xi\!\left[f^3 - \frac{1}{2}(\partial_\xi f)^2\right]
\label{KdV 2}
\eeq
and the Poisson bracket
\beq
\bigl\lbrace f(\xi),f(\xi')\bigr\rbrace = -\,\partial_\xi\delta(\xi-\xi').
\label{KdV 3}
\eeq 

The KdV equation supports an infinite number of polynomial integrals of motion,\cite{Soliton_books} all in involution with each other with respect to the Poisson bracket \eq{KdV 3}. Among these integrals of motion are the dimensionless momentum
\beq
\mathcal P = \int\!d\xi\,f^2
\label{KdV 4}
\eeq
and the Hamiltonian \eq{KdV 2} itself; the latter defines the dimensionless energy. Substituting the single-soliton solution\cite{Soliton_books} of \Eq{KdV 1},
\beq
f_0(\xi,\tau) = \frac{\,2\mathcal A^2}{\cosh^2\bigl[\mathcal A(\xi - 4\mathcal A^2\tau)\bigr]},
\label{KdV 5}
\eeq
into Eqs.~\eq{KdV 2} and \eq{KdV 4} and excluding the parameter $\mathcal A$, we obtain the relation
\beq
\mathcal H_0 = \frac{1}{5}\!\left(\frac{3\mathcal P_0}{2}\right)^{5/3}
\label{KdV 6}
\eeq
between the corresponding values of the dimensionless energy and momentum. 

In this paper, we study one-dimensional interacting quantum systems with a well-defined classical limit described by the KdV equation. As we will demonstrate in \Sec{models}, in these systems the low-energy right- and left-moving excitations decouple from each other. Focusing on this regime and on, say, the right-moving excitations, we introduce a bosonic field $\Phi$ obeying the commutation relation 
\beq
\bigl[\Phi(x),\Phi(y)\bigr] = i\pi\,\sgn(x-y)
\label{KdV 7}
\eeq
and write the Hamiltonian of the right-movers as a sum of two contributions,
\beq
H = vP + H_\text{KdV}, 
\label{KdV 8}
\eeq
where 
\beq
P = \frac{\hbar}{4\pi}\!\int_0^{L}\!dx\,\colon\!(\partial_x\Phi)^2\colon
\label{KdV 9}
\eeq
is the momentum operator and  
\beq
H_\text{KdV}
= \frac{\hbar^2}{12\pi m_\ast}\!\int_0^{L}\!dx
\,\colon\!\Bigl[(\partial_x\Phi)^3 - a_\ast(\partial^2_x\Phi)^2\Bigr]\colon
\label{KdV 10}
\eeq
represents the leading perturbation. In Eqs.~\eq{KdV 9} and~\eq{KdV 10}, $L$ is the size of the system, and the colons denote the normal ordering with respect to the bosonic vacuum. The parameters $v$, $m_\ast$ and $a_\ast$ have the units of velocity, mass and length, respectively; their origin and physical meaning will be elucidated below. We are interested in the excitations of the Hamiltonian \eq{KdV 8} with wavelengths of order $a_\ast$. Provided that the length scale~$a_\ast$ is small compared with $L$, spectra of such excitations are not sensitive to the precise form of the boundary conditions imposed on the field $\Phi(x)$, which we take to be periodic,  
\beq
\Phi(x) = \Phi(x + L).
\label{KdV 11}
\eeq 

Comparison of Eqs.~\eq{KdV 7}, \eq{KdV 9}, and \eq{KdV 10} with Eqs.~\eq{KdV 3}, \eq{KdV 4}, and \eq{KdV 2}, respectively, shows that, apart from rescaling, $\partial_x\Phi$ is merely the quantized version of $f$. The classical model defined by Eqs.~\eq{KdV 2} and \eq{KdV 3} and its quantum version\cite{Sasaki,Pogrebkov} defined by Eqs.~\eq{KdV 7} and \eq{KdV 10} represent the so-called first Hamiltonian structure of the KdV equation.\cite{Soliton_books} The quantum KdV model is believed to be integrable, with the first few integrals of motion having the same form\cite{Sasaki} as the corresponding classical expressions. These models should be distinguished from those corresponding to the second Hamiltonian structure, in which both the Hamiltonian and the Poisson bracket have forms different from Eqs.~\eq{KdV 2} and~\eq{KdV 3}.\cite{Soliton_books,Pogrebkov} Whereas quantum KdV models of the latter type have received much attention\cite{Pogrebkov,qkdv1,qkdv2,qkdv3} due to their relevance to conformal field theory,\cite{CFT} the former arise naturally as the low-energy description of one-dimensional interacting quantum systems, see, e.g., Refs.~[\onlinecite{ISG_RMP,PM_WignerToda,KA,Protopopov 2013,Protopopov 2014}].

Because the operators $P$ and $H_\text{KdV}$ commute, the excitation spectrum of the Hamiltonian \eq{KdV 8} has the form
\beq
\varepsilon(p) = vp + \varepsilon_\text{KdV}(p),
\label{KdV 12}
\eeq
where $\varepsilon$, $p$, and $\varepsilon_\text{KdV}$ are eigenvalues of the operators $H$, $P$, and $H_\text{KdV}$, respectively. In order to find the nonlinear in $p$ contribution $\varepsilon_\text{KdV}(p)$ in \Eq{KdV 12}, we consider the Heisenberg equation of motion \mbox{$\partial_t \Phi = i\hbar^{-1}\!\bigl[H_\text{KdV},\Phi\bigr]$}, which after a change of variables 
\beq
x = a_\ast \xi, 
\quad
t = t_\ast\tau,
\quad
t_\ast = \frac{3m_\ast^\pdag a_\ast^2}{\hbar\,}
\label{KdV 13}
\eeq
takes the form
\beq
\partial_\tau\Phi + \frac{3}{2}\colon\!(\partial_\xi\Phi)^2\colon \!+ \partial^3_\xi\Phi = 0.
\label{KdV 14}
\eeq 
We now differentiate \Eq{KdV 14} with respect to $\xi$ and treat the field
\beq
F = \frac{1}{2}\,\partial_\xi\Phi.
\label{KdV 15}
\eeq
as a classical variable. Neglecting quantum fluctuations of $F$ amounts to replacing $\langle \colon\! F^2\colon\!\rangle$ with $\langle F\rangle^2$ and leads to the KdV equation \eq{KdV 1} for the expectation value \mbox{$f = \langle F\rangle$}. Solution of this equation satisfying the condition 
\mbox{$\int_0^{L/a_\ast}\!d\xi\, f(\xi,\tau) = 0$} [see Eqs.~\eq{KdV 11} and \eq{KdV 15}] can be written as
\beq
f(\xi,\tau) = -\,\mathcal N_0 \frac{\,a_\ast}{L\,} + f_0(\xi,\tau),
\label{KdV 16}
\eeq 
where \mbox{$\mathcal N_0 = \int\!d\xi\,f_0(\xi,\tau) = 2(3\mathcal P_0/2)^{1/3}$} is the dimensionless mass of the soliton\cite{Soliton_books} and $f_0$ is given by \Eq{KdV 5}. 
We now take the expectation values of $P$ and $H_\text{KdV}$ [see Eqs.~\eq{KdV 9} and \eq{KdV 10}], neglect quantum fluctuations, and compare the resulting expressions with Eqs.~\eq{KdV 2} and~\eq{KdV 4}. In the limit $a_\ast/L\to 0$, this yields
\beq
p = \frac{\hbar_\past}{\pi a_\ast}\mathcal P_0,
\quad
\varepsilon_\text{KdV} = \frac{2\hbar_\past}{\pi t_\ast} \mathcal H_0.
\label{KdV 17}
\eeq
Upon introducing the momentum and energy scales
\beq
p_\ast = \frac{\,3\hbar_\past}{\,2a_\ast},
\quad
\varepsilon_\ast = \frac{27}{8}\frac{\hbar_\ast}{t_\ast} = \frac{p_\ast^2\,}{2m_\ast},
\label{KdV 18}
\eeq
we write the KdV contribution to the spectrum as 
\beq
\varepsilon_\text{KdV}(p) =\varepsilon_\ast e(p/p_\ast).
\label{KdV 19}
\eeq 
For the soliton excitation the dimensionless function $e(s)$ in \Eq{KdV 19} is given by
\beq
e_\text{soliton}(s) = \frac{3}{5}\left(\frac{2\pi}{3}\right)^{2/3} s^{5/3},
\label{KdV 20}
\eeq
see Eqs.~\eq{KdV 6} and \eq{KdV 17}.

In addition to the solitons, the KdV equation \eq{KdV 1} has delocalized solutions describing periodic waves, the cnoidal waves.\cite{KdV_original} The energy \eq{KdV 2} and momentum \eq{KdV 4} associated with such solutions diverge in the limit of infinite system size unless the waves have vanishingly small amplitude. Accordingly, periodic wave solutions of interest here correspond to the harmonic regime when the nonlinear term in \Eq{KdV 1} can be neglected. This leads to the relation $\Omega = -\,Q^3$ between the dimensionless frequency~$\Omega$ and the wave number $Q$. 
In classical mechanics the energy and momentum of such a wave are proportional to the square of its amplitude. In quantum mechanics, however, the periodic wave solution corresponds to a phonon with well defined energy and momentum. Restoring the units, we obtain \mbox{$\varepsilon_\text{KdV} = \hbar_{}\Omega/t_\ast$} and \mbox{$p = \hbar Q/a_\ast$}. The wave dispersion relation can now be converted to the phonon spectrum. It has the form of \Eq{KdV 19} with~$e(s)$ given by
\beq
e_\text{phonon}(s) = -\,s^3.
\label{KdV 21}
\eeq

Use of the classical equation of motion for the evaluation of the excitation spectrum relies on the assumption that the quantum uncertainty of the field $F$ is negligible. With the help of Eqs.~\eq{KdV 7} and \eq{KdV 15}, the magnitude of quantum fluctuations of $F$ with the length scale $\Delta_\xi \sim 1/\mathcal A$ relevant for a classical soliton [see \Eq{KdV 5}] can be estimated as \mbox{$\delta F\sim 1/\Delta_\xi\sim \mathcal A$}. The condition of applicability of the classical description \mbox{$\delta F\ll \langle F\rangle\sim \mathcal A^2$} then leads to $\mathcal A\gg 1$, which translates to $\mathcal P_0\gg 1$ for the dimensionless classical momentum \eq{KdV 4} and to $p\gg p_\ast$ in \Eq{KdV 19} or $s\gg 1$ in Eqs.~\eq{KdV 20} and \eq{KdV 21}. 
 
For excitations with small momenta $p\lesssim p_\ast$ the quantum fluctuations can no longer be neglected, and Eqs.~\eq{KdV 20} and \eq{KdV 21} are inapplicable. Fortunately, at \mbox{$p\to 0$}, i.e., deep in the quantum regime, the KdV Hamiltonian \eq{KdV 10} allows  further simplification. Indeed, the second term on the right-hand side of \eq{KdV 10} has a higher scaling dimension than the first one, hence its effect on long-wavelength excitations reduces to being merely a small perturbation.\cite{Rozhkov} Neglecting this term and employing the well-known mapping\cite{Pogrebkov,Mattis,Haldane_LL} between bosons and fermions, 
\beq
\Psi(x) = \frac{\,1}{\sqrt{L}}\,\colon\!e^{i\Phi(x)\,}\!\colon\!,
\label{KdV 22}
\eeq
we arrive at the fixed-point Hamiltonian\cite{Rozhkov} 
\beq
H_\text{fermion} = \frac{\hbar^2}{2m_\ast}\!\int_0^L\!dx\,
\triplecolon\,(\partial_x\Psi)^\dagger (\partial _x\Psi)\,\triplecolon\,,
\label{KdV 23}
\eeq
where the symbols $\triplecolon$ denote the normal ordering with respect to the fermionic vacuum in which all single-particle states with positive (negative) wave numbers are  empty (occupied).\cite{Pogrebkov,Mattis,Haldane_LL} Application of the identity \eq{KdV 22} also yields the fermionic representation\cite{Pogrebkov,Mattis,Haldane_LL} of the momentum operator~\eq{KdV 9},
\beq
P = \!-\,i\hbar\!\int_0^L\!dx\,
\triplecolon\,\Psi^\dagger(x)\,\partial_x\Psi(x)\,\triplecolon\,,
\label{KdV 24}
\eeq
and the relation\cite{Mattis,Haldane_LL} 
\beq
\partial_x\Phi = 2\pi\, \triplecolon\,\Psi^\dagger(x)\Psi(x)\,\triplecolon\,,
\label{KdV 25}
\eeq
which shows that the field $F$ introduced in \Eq{KdV 15} above is proportional to the density of the effective fermions. 

The boundary condition \eq{KdV 11} and the relation \eq{KdV 25} imply that $\delta N = \int_0^L\!dx\,\triplecolon\,\Psi^\dagger(x)\Psi(x)\,\triplecolon = 0$. Any eigenstate of $\delta N$ with $\delta N = 0$ can be viewed as a superposition of the particle- and hole-type elementary excitations. The particle excitation is obtained by promoting a fermion from the Fermi level (which corresponds to the single-particle state with wave number zero) to one of the unoccupied single-particle states, whereas in the hole excitation a fermion is removed from one of the occupied single-particle states and placed at the Fermi level. It is easy to see that such particle and hole excitations are eigenstates of $H_\text{fermion}$ and~$P$. The corresponding eigenvalues $\varepsilon_\text{fermion}$ and~$p$ obviously satisfy
\beq
\varepsilon_\text{fermion}(p) = \pm\,\frac{p^2}{2m_\ast},
\label{KdV 26}
\eeq 
where the $+(-)$ sign corresponds to the particle (hole) excitation.  This expression can be cast in the form~\eq{KdV 19} with 
\beq
e_\text{fermion}(s) = \pm\,s^2.
\label{KdV 27}
\eeq  

Note that Eqs.~\eq{KdV 20}, \eq{KdV 21}, and \eq{KdV 27} yield \mbox{$|e(s)|\sim 1$} at \mbox{$s\sim 1$}. This observation indicates that the classical-to-quantum crossover at \mbox{$p\sim p_\ast$} is the only crossover that takes place in the system. 

In view of Eqs.~\eq{KdV 15}, \eq{KdV 16} and \eq{KdV 25}, the classical soliton excitation corresponds to a hump in the fermionic density made up of \mbox{$\mathcal N\sim \mathcal N_0\sim s^{1/3}$} fermions drawn from a uniform background. According to the discussion above, the classical description is applicable as long as $\mathcal N\gg 1$. On the contrary, the quantum counterpart of the classical soliton, the particle excitation, has exactly one excited fermion, which can be interpreted as~\mbox{$\mathcal N = 1$}. Quite naturally, the classical-to-quantum crossover occurs at~$\mathcal N\sim 1$, i.e., when the discreteness of $\mathcal N$ can no longer be ignored.

In the above consideration we treated $a_\ast$ in \Eq{KdV 10} as a positive parameter. In fact, it can have either sign. However, it is easy to see that the operators $H_\text{KdV}(a_\ast)$ and $H_\text{KdV}(-\,a_\ast)$ are related by the transformation \mbox{$\Phi\to-\,\Phi$}, which amounts to the particle-hole transformation \mbox{$\Psi\to\Psi^\dagger$} for fermions, see \Eq{KdV 22}. Under such particle-hole transformation \mbox{$H_\text{KdV}(-a_\ast)\to-\,H_\text{KdV}(a_\ast)$} and $P\to P$. Accordingly, the spectra of the elementary excitations of $H_\text{KdV}(-a_\ast)$ are related to those of $H_\text{KdV}(a_\ast)$ and can be written in the form similar to~\Eq{KdV 19} as
\beq
\varepsilon_\text{KdV}(p) = - \,\varepsilon_\ast e(p/p_\ast).
\label{KdV 28}
\eeq
Note that the particle-hole transformation changes the sign of the fermionic density \eq{KdV 25}. Thus, whereas the solitonic excitation of $H_\text{KdV}(a_\ast)$ carries $\mathcal N > 0$ fermions and corresponds to a hump in the fermionic density, its particle-hole-transformed twin has \mbox{$\mathcal N < 0$}, which amounts to a depression. In nonlinear optics literature these two kinds of solitons are often referred to as bright and dark solitons, respectively.\cite{BrightDark}
  
\section{Models}
\label{models}

Instead of attempting to evaluate the excitation spectrum of the quantum KdV model \eq{KdV 10} directly, we rely on well-known solvable models, viz.~the Lieb-Liniger model and the quantum Toda model. In this section, we demonstrate that in judiciously chosen scaling limits the low-energy theories describing these models reduce to that defined by Eqs.~\eq{KdV 7}-\eq{KdV 10}. The reduction hinges on the smallness of the parameter 
\beq
\zeta = \frac{\varepsilon_\ast}{v p_\ast} = \frac{\,p_\ast}{2m_\ast v\,},
\label{M 1}
\eeq
which characterizes the relative magnitude of the KdV contribution to the excitation spectra [see \Eq{KdV 12}] evaluated near the classical-to-quantum crossover at~\mbox{$p\sim p_\ast$}. Although the low-energy Hamiltonians for both the Lieb-Liniger and the quantum Toda models contain non-universal contributions absent in \Eq{KdV 8}, below we show that these contributions affect the spectra only in the second order in $\zeta$, and
\beq
\varepsilon(p) = vp_\ast\!\left[s \pm \zeta e(s) + O\bigl(\zeta^2\bigr)\right],
\quad
s = p/p_\ast.
\label{M 2}
\eeq
Here the $+(-)$ sign corresponds to the quantum Toda (Lieb-Liniger) model, and $e(s)$ [see \Eq{KdV 19}] are the dimensionless crossover functions describing the spectra of the quantum KdV model \eq{KdV 10}.

\subsection{Lieb-Liniger model}
\label{LiebLiniger}

The Lieb-Liniger model\cite{Lieb,KBI,Yang}
\beq
H = \frac{\,\hbar^2}{2m\,}\!\left\lbrace
-\sum_{l}\frac{\,\partial^2}{\partial x_l^2} 
+ \sum_{\,l\,\neq\, l'}\!c\,\delta(x_l - x_{l'}\!)
\right\rbrace
\label{LL 1}
\eeq
describes bosons with contact interaction. We are interested in the thermodynamic limit when both the number of particles $N$ and the system size $L$ are taken to infinity with their ratio, the mean density $n_0 = N/L$, kept fixed. The interaction strength is characterized by the dimensionless parameter\cite{Lieb} $\gamma = c/n_0$. In the weak repulsion regime considered here, $0 < \gamma\ll 1$. The Hamiltonian \eq{LL 1} can also be written in the second-quantized form
\beq
H =  \frac{\,\hbar^2}{2m\,}\!\int\!dx\,
\Bigl[(\partial_x\psi)^\dagger (\partial _x\psi) + c_{}n^2(x)\Bigr],
\label{LL 2}
\eeq
where $n(x) = \psi^\dagger(x)\psi(x)$ is the density operator. Hereinafter, products of quantum fields at the same spatial point, such as $n^2(x)$ in \Eq{LL 2}, are to be understood as being normal-ordered with respect to the appropriate vacuum states, cf. \Sec{KdV}.

We are interested in excitations with wavelengths much longer than the mean interparticle distance $1/n_0$. Following the standard prescription,\cite{Haldane,Popov} we write
\beq
\psi(x) = \sqrt{n(x)\,}e^{-i\vartheta(x)}.
\label{LL 3}
\eeq
Here $n(x)$ is the coarse-grained (averaged over a region much larger than $1/n_0$) particle density, which can be regarded as a continuous function of $x$, and the field $\vartheta$ satisfies\cite{Haldane,Popov} 
\beq
[n(x),\vartheta(y)] = -\,i\delta(x-y).
\label{LL 4}
\eeq 
We will assume that $\vartheta$ obeys the periodic boundary condition \mbox{$\vartheta(x + L) = \vartheta(x)$}. This assumption amounts\cite{Haldane} to restricting one's attention to excitations near the zero-momentum ground state. 

Substitution of $\psi(x)$ in the form \eq{LL 3} into \Eq{LL 2} yields
\beq
H = \frac{\,\hbar^2}{2m\,}\!\int\!dx\!
\left[n(\partial_x\vartheta)^2 + \frac{1}{4n}(\partial_x n)^2 + c_{}n^2\right].
\label{LL 5}
\eeq
For low-energy excitations, deviations of the density $n(x)$ from its mean value $n_0$ are small. It is therefore convenient to write $n(x)$ as 
\beq
n(x) = n_0 + \frac{1}{\pi}\partial_x\varphi,
\label{LL 6}
\eeq
where the new bosonic field $\varphi$ satisfies
\beq
[\partial_x\varphi,\vartheta(y)] = -\,i\pi\delta(x-y)
\label{LL 7}
\eeq 
and obeys the periodic boundary condition. Successive approximations for the low-energy Hamiltonian are obtained by substituting \Eq{LL 6} into \Eq{LL 5} and expanding in powers of $\partial_x\varphi$.

The leading contribution to the low-energy Hamiltonian contains operators of scaling dimension two and it has the standard Luttinger liquid form\cite{Haldane_LL,Haldane,EL,Popov}
\beq
H_0  = \frac{\hbar v}{2\pi}\!\int\!dx\bigl[K(\partial_x\vartheta)^2 + K^{-1}(\partial_x\varphi)^2\bigr].
\label{LL 8}
\eeq
Here 
\beq
v = \frac{\hbar n_0}{m\,} \sqrt{\gamma\,},
\quad
K = \frac{\pi\hbar n_0}{mv\,} 
= \frac{\,\pi}{\sqrt{\gamma\,}}
\label{LL 9}
\eeq
are the sound velocity and the Luttinger liquid parameter evaluated in the leading order in $\gamma\ll 1$. Instead of $\varphi$ and~$\vartheta$, it is convenient to introduce the right- and left-moving fields
\beq
\varphi_\pm(x) = \frac{\,1}{\sqrt{K\,}}\,\varphi(x) \mp \sqrt{K\,}\vartheta(x),
\label{LL 10}
\eeq
which satisfy 
$\bigl[\partial_x\varphi_+,\varphi_-(y)\bigr] = \bigl[\varphi_+(x),\partial_y\varphi_-\bigr] = 0$ and
\beq
\bigl[\varphi_\pm(x),\varphi_\pm(y)\bigr] = \pm\,i\pi\,\sgn(x-y).
\label{LL 11}
\eeq
Writing the Luttinger liquid Hamiltonian \eq{LL 8} in terms of $\varphi_\pm$ reveals its chiral (i.e., diagonal in the basis of the right- and left-movers) structure,
\beq
H_0 = \frac{\hbar v}{4\pi}\!\int\!dx\!\sum_{\nu\,=\,\pm}\!(\partial_x\varphi_\nu)^2.
\label{LL 12}
\eeq
 
In addition to the Hamiltonian, we need the momentum operator $P = -i\hbar\!\int\!dx\,\psi^\dagger\partial_x\psi$. Substituting here \Eq{LL 3} and taking into account \Eq{LL 6}, we obtain
\beq
P = -\,\hbar\!\int\!dx\,n_{}\partial_x\vartheta = -\,\frac{\hbar}{\pi}\!\int\!dx\,(\partial_x\varphi)(\partial_x\vartheta).
\label{LL 13}
\eeq
Written in terms of the right- and left-movers $\varphi_\pm$, see \Eq{LL 10}, the momentum \eq{LL 13} takes the form
\beq
P = P_+ + P_-,
\quad
P_\pm = \pm\,\frac{\hbar}{4\pi}\!\int\!dx\,(\partial_x\varphi_\pm)^2.
\label{LL 14}
\eeq
Comparison of Eqs.~\eq{LL 12} and \eq{LL 14} shows that the Luttinger liquid Hamiltonian \eq{LL 12} can be written as 
\beq
H_0= v(P_+ - P_-).
\label{LL 15}
\eeq 

The nonlinear corrections to the excitation spectra arise due to higher-order terms in the gradient expansion of the Hamiltonian \eq{LL 5}. Collecting the chiral terms with scaling dimensions three and four in this expansion, we write the resulting contribution in the form of the quantum KdV Hamiltonian \eq{KdV 10},
\beq
H_1 = \frac{\hbar^2}{12\pi m_\ast}\!\int\!dx\sum_\nu\,
\Bigl[(\partial_x\varphi_\nu)^3 + a_\ast(\partial^2_x\varphi_\nu)^2\Bigr],
\label{LL 16}
\eeq
where the effective mass $m_\ast$ and the emergent length scale $a_\ast$ are defined by
\beq
\frac{\,m_\past}{\,m_\ast} = \frac{3}{4\sqrt{K}},
\quad
{a_\ast n_0} = \frac{\,K^{3/2}}{\,2\pi^2\,} 
\label{LL 17}
\eeq
and satisfy $m/m_\ast \ll 1$, $a_\ast n_0\gg 1$. 
Note that the result for the effective mass agrees with the general expression\cite{Pereira,ISG_RMP}
\beq
\frac{\,m_\past}{\,m_*}  = \frac{1}{2v\sqrt{K}}\frac{d (vn_0)}{dn_0}, 
\label{LL 18}
\eeq
valid for any Galilean-invariant system.

Although the second term in \Eq{LL 16} has a higher scaling dimension than the first one, its effect on the excitations with wavelengths of order $a_\ast$  [or, equivalently, with momenta of order $p_\ast\sim \hbar/a_\ast$, see \Eq{KdV 18}] is comparable with that of the first term in \Eq{LL 16}. The key observation is that for weakly interacting bosons 
\beq
\frac{\,p_\ast}{\hbar n_0} = \frac{3\pi}{\,K^{3/2}}\ll 1,
\label{LL 19}
\eeq
i.e., such excitations belong to the realm of the long-wavelength description based on the gradient expansion. Changing the integration variable in Eqs.~\eq{LL 12} and \eq{LL 16} to $\xi = x/a_\ast$, we write the expansion \mbox{$H = H_0 + H_1 +\ldots$} as
\beq
H = vp_\ast\bigl( h_0 + \zeta h_1 + \zeta h_1^\prime + \zeta^2 h_2 +\ldots\bigr),
\label{LL 20}
\eeq
with the parameter $\zeta$ introduced in \Eq{M 1}. Using Eqs.~\eq{LL 9}, \eq{LL 17}, and \eq{LL 19}, we find
\beq
\zeta = \frac{9\,}{8K} \ll 1.
\label{LL 21}
\eeq
The operators $h_0$ and $h_1$ in \Eq{LL 20} are given by
\begin{subequations}
\beqa
h_0 &=& \frac{1}{6\pi}\!\int\!d\xi\sum_{\nu}\,(\partial_\xi\varphi_\nu)^2,
\label{LL 22a} \\
h_1  &=& \frac{2}{27\pi}\!
\int\!d\xi\sum_{\nu}\,\Bigl[(\partial_\xi\varphi_\nu)^3 + (\partial^2_\xi\varphi_\nu)^2\Bigr],
\qquad
\label{LL 22b}
\eeqa
\end{subequations}
cf.~Eqs.~\eq{LL 12} and \eq{LL 16}. The third term in \Eq{LL 20} includes non-chiral contributions with scaling dimensions three and four omitted in \Eq{LL 22b}, $(\partial_\xi\varphi_{\pm})^2(\partial_\xi\varphi_{\mp})$ and $(\partial^2_\xi\varphi_{+})(\partial^2_\xi\varphi_{-})$. Contributions in \Eq{LL 20} that are higher order in $\zeta$ originate in the expansion of the second term in \Eq{LL 5}, the so-called quantum pressure,\cite{Pitaevskii} in $\partial_x \varphi$. The first term in this series yields $h_2$ in \Eq{LL 20}, which contains operators of scaling dimension five $(\partial_\xi^2\varphi_\nu)^2(\partial_\xi\varphi_\nu)$.

The expansion \eq{LL 20} allows us to classify various terms in the low-energy Hamiltonian according to the order of magnitude of their contributions to the energy of chiral excitations. It should be emphasized that this classification is different from the usual notion of scaling dimension. Indeed, the latter is relevant for the description of excitations in the limit of infinitely long wavelengths, whereas we are interested in excitations characterized by long but finite wavelengths of order $a_\ast$.

We proceed by singling out the first two terms in the expansion \eq{LL 20},
$\widetilde H = vp_\ast\bigl( h_0 + \zeta h_1\bigr)$, treating the remainder of the expansion 
as a perturbation. Taking into account that $\widetilde H$ commutes with $P_\pm$, we consider a simultaneous eigenstate of the operators $P_+$, $P_-$, and $\widetilde H$ with eigenvalues  $p_+ = p\sim p_\ast$, $p_- = 0$, and $\widetilde\varepsilon$, respectively. In the right-moving sector (i.e., when acting on states with $p_- = 0$, such as the one we discuss), the unperturbed Hamiltonian $\widetilde H$ coincides with that defined by Eqs.~\eq{KdV 7}-\eq{KdV 10}, with $a_\ast$ in $H_\text{KdV}$ [see \Eq{KdV 10}] replaced with $-\,a_\ast$. Using Eqs.~\eq{KdV 12}, \eq{KdV 28}, and \eq{M 1}, we obtain $\widetilde\varepsilon(p) = vp_\ast\bigl[s - \zeta e(s)\bigr]$ with $s = p/p_\ast$. 

Perturbation theory in $H - \widetilde H = vp_\ast\bigl(\zeta h_1^\prime + \zeta^2 h_2 +\ldots\bigr)$ establishes the correspondence between the eigenstate of~$\widetilde H$ we consider and the eigenstate of the full Hamiltonian $H$ with the energy $\varepsilon = \widetilde\varepsilon + \delta\varepsilon$. The latter state is also an eigenstate of the total momentum $P$ [see \Eq{LL 14}] with the eigenvalue $p$. The leading corrections to the energy arise in the second order in~$h_1^\prime$ (note that $h_1^\prime$ has zero expectation value in the right-moving eigenstate of $\widetilde H$), and in the first order in $h_2$. Both contributions can be estimated as $\delta\varepsilon\sim v p_\ast\zeta^2$. Accordingly, $\varepsilon(p)  = \widetilde\varepsilon(p) \,+\, O\bigl(vp_\ast\zeta^2\bigr)$, leading to \Eq{M 2}.  

\subsection{Quantum Toda model}
\label{QuantumToda}

The classical Toda model\cite{Toda} describes a chain of $N$ particles with exponential nearest-neighbor interaction,
\beq
H = \sum_l\left[\frac{\,p_l^2}{2m\,} + V_0 e^{-\,2D_l/a_0} \right],
\quad
D_l = x_{l+1} - x_l.
\label{T 1}
\eeq
We consider a system of size $L$ in the thermodynamic limit taken at a fixed particle density $n_0 = N/L$. The potential energy in \Eq{T 1} is minimized in the ordered configuration 
$x_1 < x_2 < \cdots < x_{N}$, in which the distance between neighboring particles $D_l$ equals its mean value~$1/n_0$. At finite but sufficiently low energies, the deviations from the mean $\delta D_l = D_l - 1/n_0$ remain small compared with the interaction range $a_0$. In such a weakly anharmonic regime the interaction potential in \Eq{T 1} can be expanded in series in $\delta D_l/a_0$.  After taking a continuum limit and making an appropriate change of variables, the leading terms of such expansion can be cast\cite{Toda} in the form of the KdV Hamiltonian~\eq{KdV 2}. The model \eq{T 1} is integrable and supports solitonic excitations with arbitrarily high energies, well beyond the weakly anharmonic regime.\cite{Toda} However, these excitations, the Toda solitons, can no longer be described in the KdV framework.

In the quantum Toda model,\cite{Gutzwiller,Sklyanin,Sutherland_book,Sutherland}  coordinates  of the particles $x_l$ and their momenta $p_l$ are replaced with the operators satisfying 
$[x_l,p_{l'}] = i\hbar_{}\delta_{l,l'}$. Unlike the classical Toda model, the very existence of the weakly anharmonic regime is not \textit{a priori} guaranteed even at low energies but hinges on the smallness of quantum fluctuations of~$\delta D_l$. The lower bound $\delta D$ on the magnitude of these fluctuations can be estimated as the amplitude of zero-point oscillations of the positions of the particles near the respective potential minima. It is convenient to characterize the range and the strength of the interaction potential in \Eq{T 1} by the dimensionless parameters
\beq
\alpha = \frac{1}{\,a_0 n_0},
\quad
\beta = \frac{1}{\,2\alpha e^{\alpha}}\frac{\sqrt{m V_0}\,}{\,\,\hbar n_0} . 
\label{T 2}
\eeq
In terms of these parameters, $\delta D\sim a_0 \beta^{-1/2}$ and the necessary condition for the existence of the weakly anharmonic regime $\delta D \ll a_0$ translates to 
\beq
\beta \gg 1,
\label{T 3}
\eeq
irrespective of the value of $\alpha$.

The quantum Toda model is integrable,\cite{Gutzwiller,Sklyanin} and its properties can be studied analytically\cite{Sklyanin} at arbitrary values of the parameters $\alpha$ and $\beta$. However, our goal is to describe excitations in the universal KdV regime rather than to explore various regimes of the quantum Toda model. Therefore, we make a simplifying assumption
\beq
\alpha \gg 1.
\label{T 4}
\eeq
The advantage of working in the dilute limit \eq{T 4} is two-fold. First, as shown below, in this limit $\zeta\sim 1/\beta$, hence \Eq{T 3} guarantees the existence of the KdV regime. Second, excitations of the Toda model in the dilute limit can be studied\cite{Sutherland_book,Sutherland}  by considering instead the hyperbolic Calogero-Sutherland model\cite{Calogero,Sutherland_book,Sutherland} 
\beq
H = \frac{\,\hbar^2}{2m\,}\!\left\lbrace
-\sum_{l}\frac{\,\partial^2}{\partial x_l^2} 
+\sum_{\,l\,\neq\,l'}
\frac{\lambda(\lambda - 1)}{a_0^2\sinh^2\bigl[(x_l - x_{l'})/a_0\bigr]}
\right\rbrace.
\label{T 5} 
\eeq
For large $\lambda$ the distance between neighboring particles is close to $1/n_0 \gg a_0$, see \Eq{T 4}. Therefore, the sinh function in \Eq{T 5} can be approximated by exponential, and the interaction can be restricted to nearest neighbors. Thus, the model \eq{T 5} is equivalent\cite{Sutherland,Sutherland_book} to the Toda model \eq{T 1} with 
\beq
V_0 = \frac{4\hbar^2\lambda(\lambda - 1)}{m a_0^2}.
\label{T 6}
\eeq 
Substitution into \Eq{T 2} gives 
\beq
\beta = \sqrt{\lambda(\lambda - 1)} \,e^{-\alpha} \approx \lambda e^{-\alpha},
\label{T 7}
\eeq 
and the condition \Eq{T 3} yields 
\beq
\lambda \gg e^{\alpha}.
\label{T 8}
\eeq
The inequalities \eq{T 4} and \eq{T 8} define the Toda limit of the hyperbolic Calogero-Sutherland model \eq{T 5}. Unlike the quantum Toda model \eq{T 1}, the hyperbolic Calogero-Sutherland model \eq{T 5} is solvable by the asymptotic Bethe ansatz,\cite{Sutherland_book,Sutherland} which allows it to be handled on equal footing with the Lieb-Liniger model.

For long-wavelength excitations, $p_l$ and $D_l$ in \Eq{T 1} vary with $l$ on the scale much larger than unity, and $l$ can be treated as a continuous variable. Replacing the summation over $l$ in \Eq{T 1} with the integration, we write the low-energy Hamiltonian as
\beq
H = \int\!dl \left[\frac{\,p^2(l)}{2m} + V_0 e^{-2D(l)/a_0}\right],
\label{T 9}
\eeq
where $D(l) =  x(l+1) - x(l)$, and the fields $x(l)$ and $p(l)$ satisfy $[x(l),p(l')] = i\hbar_{}\delta(l-l')$. 

The Hamiltonian \eq{T 9} describes the strongly interacting quantum fluid in terms of the so-called Lagrangian variables,\cite{Landau,MA,PM_WignerToda} in which the position of the fluid element is specified by $l$ rather than by its physical coordinate~$x(l)$. The consideration in Secs.~\ref{KdV} and~\ref{LiebLiniger}, however, follows the conventional bosonization scheme\cite{Haldane,Popov,EL} based on the Eulerian\cite{Landau,MA} description of the quantum fluid. A switch to the Eulerian formulation is accomplished using 
\beq
dl = n(x) dx,
\label{T 10}
\eeq
where $n(x)$ is the coarse-grained density operator, and writing the momentum per particle $p(l)$ as\cite{MA}
\beq
p(l) = -\hbar_{}\partial_x\vartheta,
\label{T 11}
\eeq
where the field $\vartheta$ has essentially the same meaning as that in \Eq{LL 3} and satisfies the same commutation relation~\eq{LL 4}. Indeed, substituting Eqs.~\eq{T 10} and \eq{T 11} into the expression $P = \int\!dl\,p(l)$, we recover \Eq{LL 13} for the momentum operator, $P = -\hbar\!\int\!dx\,n(x) \partial_x\vartheta$.

The Eulerian form of the interparticle distance $D(l)$ in \Eq{T 9} follows from the identity $D(l) = \bigl[e^{\partial_l} - 1\bigr]x(l)$. Replacing $\partial_l$ with $n^{-1}\partial_x$ here [see \Eq{T 10}], we find
\beqa
D &=& \big(e^{n^{-1}\partial_x} - 1\bigr)x 
\nn \\
&=& \frac{1}{n} + \frac{1}{2!}\, \frac{1}{n}\partial_x\frac{1}{n} + \frac{1}{3!}\, \frac{1}{n}\partial_x\frac{1}{n} \partial_x\frac{1}{n} + \ldots,
\quad
\label{T 12}
\eeqa
i.e., $D$ depends on the density $n(x)$ and its derivatives. Predictably, $D$ reduces to $1/n$ in the long-wavelength limit. 

Using Eqs.~\eq{T 10} and \eq{T 11}, we rewrite the Hamiltonian~\eq{T 9} as
\beq
H = \frac{\,\hbar^2}{2m\,}\!\int\!dx\,n(\partial_x\vartheta)^2 + U[n],
\label{T 13}
\eeq
where the interaction energy $U[n]$ is a functional of density,
\beq
U[n] = V_0\!\int\!dx\,n e^{-2D/a_0}
\label{T 14}
\eeq
with $D$ given by \Eq{T 12}. Note that the field $\vartheta$ enters the Hamiltonian \eq{T 13} only via the kinetic energy term, which has the same form as that in \Eq{LL 5}. This is a direct consequence of the Galilean invariance of the models \eq{LL 1} and \eq{T 1}.

The remaining consideration parallels that in \Sec{LiebLiniger}: we write the density $n(x)$ in the form \eq{LL 6} and expand the Hamiltonian \eq{T 13} in $\partial_x\varphi$. The leading contribution in this expansion is the Luttinger liquid Hamiltonian \eq{LL 8} with 
\beq
K = \frac{\pi}{4 \alpha^2\beta} \ll 1.
\label{T 15}
\eeq
In the next step, we pick terms containing integrals of $(\partial_x\varphi)(\partial_x\vartheta)^2$, $(\partial_x^2\varphi)^2$, and $(\partial_x\varphi)^3$. Expressing $\varphi$ and $\vartheta$ via the right/left-moving fields $\varphi_\pm$ [see \Eq{LL 10}] and omitting all non-chiral terms, we obtain the KdV contribution
\beq
H_1 = \frac{\hbar^2}{12\pi m_\ast}\!\int\!dx\sum_\nu\,
\Bigl[(\partial_x\varphi_\nu)^3 - a_\ast(\partial^2_x\varphi_\nu)^2\Bigr].
\label{T 16}
\eeq
Except for the sign of the second term on the right-hand side, \Eq{T 16} has the same form as \Eq{LL 16}. The effective mass $m_\ast$ and the emergent length scale $a_\ast$ in \Eq{T 16} are defined by
\begin{subequations}
\beqa
\frac{\,m\past}{m_\ast} 
&=& \alpha^2\sqrt{\beta/\pi} = \frac{\alpha}{2\sqrt{K\,}},
\label{T 17a} \\
a_\ast n_0 &=& \frac{1}{2}\sqrt{\pi\beta\,}
= \frac{\pi}{4\alpha\sqrt{K\,}},
\label{T 17b}
\eeqa
\end{subequations}
again in agreement with \Eq{LL 18}. Taking into account Eqs.~\eq{T 3} and \eq{T 4}, we find  
$m/m_\ast \gg 1$, $a_\ast n_0\gg 1$. 
Substitution of \Eq{T 17b} into \Eq{KdV 18} yields 
\beq
\frac{p_\ast}{\hbar n_0} 
= \frac{\,3}{\sqrt{\pi\beta\,}\,}
= \frac{6\alpha}{\pi}\sqrt{K\,} \ll 1.
\label{T 18}
\eeq
for the crossover momentum. The expansion parameter~$\zeta$ [see \Eq{M 1}] is then given by
\beq
\zeta = \frac{3}{8\pi\beta} = \frac{3\,}{2\pi^2}\,\alpha^2 K \ll 1.
\label{T 19}
\eeq 

Focusing on excitations with wavelengths of order $a_\ast$, we change the integration variable to $\xi = x/a_\ast$ and write the gradient expansion of low-energy Hamiltonian in the form \eq{LL 20}. The Luttinger liquid contribution in this expansion, $h_0$, is given by \Eq{LL 22a}, and the KdV contribution $h_1$ differs from \Eq{LL 22b} only in the sign of the second term on the right-hand side.
The remaining terms in the expansion \eq{LL 20} are non-universal. The third term, $h_1'$, contains non-chiral contributions with scaling dimensions three and four, 
whereas the fourth term, $h_2$, is composed of the chiral contributions $(\partial_\xi\varphi_\nu)^4$, $(\partial_\xi\varphi_\nu)^2(\partial_\xi^3\varphi_\nu)$, and $(\partial_\xi^3\varphi_\nu)^2$, which originate in the expansion of the interaction energy \Eq{T 14}. Importantly, various operators enter $h_1^\prime$ and $h_2$ with numerical coefficients that have finite limits of order unity at~$\alpha\to\infty$.

Repeating the analysis of \Sec{LiebLiniger}, we conclude that $h_1^\prime$ and $h_2$ contribute to the energy of chiral excitations with momenta $p\sim p_\ast$ only in the second order in $\zeta$. Therefore, to first order in $\zeta$ the excitation spectrum is given by Eqs.~\eq{KdV 12} and \eq{KdV 19}, leading to \Eq{M 2} with a positive sign of the second term on the right-hand side.

\section{Elementary excitations}
\label{Excitations}

As shown in Secs.~\ref{KdV} and \ref{models}, excitation spectra of the chiral model defined by Eqs.~\eq{KdV 7}-\eq{KdV 10}, the Lieb-Liniger model~\eq{LL 1}, and the quantum Toda model \eq{T 1} are given by \Eq{M 2}. The non-trivial parts of the spectra are described by the universal dimensionless crossover functions $e(s)$, with asymptotes given by Eqs.~\eq{KdV 20} and~\eq{KdV 21} in the classical limit $s\gg 1$, and by \Eq{KdV 27} in the quantum limit $s\ll 1$. In this Section we take advantage of the integrability of the Lieb-Liniger and the hyperbolic Calogero-Sutherland models. We evaluate their spectra to first order in $\zeta$ and extract the crossover functions $e(s)$.

\subsection{Excitation spectra from Bethe ansatz}
\label{Bethe}

The Lieb-Liniger model \eq{LL 1} and the hyperbolic Calogero-Sutherland model \eq{T 5} are integrable by Bethe ansatz~[\onlinecite{Lieb,Yang,KBI,Sutherland_book}]: their many-body eigenstates are parametrized by sets of $N$ rapidities $q_1, q_2,\ldots, q_{N}$, which are similar to the wave numbers of free fermions. Lieb's type I and type II excitations\cite{Lieb} can be viewed as, respectively, particle- and hole-type excitations of the corresponding Fermi sea.\cite{Lieb,Yang,KBI,Sutherland_book} Their momenta and energies are given parametrically by
\beq
p(q) = 2\pi\hbar \left|\int_{q_0}^q\!dq'\rho(q')\right|,
\quad
\varepsilon(q) = \left|\int_{q_0}^q\!dq'\sigma(q')\right|,
\label{BA 1}
\eeq
where $q > q_0~(q < q_0)$ for the type I (type II) excitations, and $q_0$ is the Fermi rapidity. The function \mbox{$\rho(q) = L^{-1}\!\sum_{i=1}^{N} \delta(q-q_i)$} is the density of rapidities in the ground state. In the thermodynamic limit, $\rho(q)$ can be viewed as a continuous function of $q$, normalized as\cite{Lieb,KBI,Sutherland_book}
\beq
\int_{-q_0}^{q_0}\!\! dq\,\rho(q) = n_0,
\label{BA 2} 
\eeq
and satisfying the Lieb equation\cite{Lieb,KBI,Sutherland_book}
\beq
\rho(q) + \frac{1}{2\pi}\!\int_{-q_0}^{q_0}\!\!dq'\Theta'(q-q')\rho(q') = \frac{1}{2\pi},
\label{BA 3} 
\eeq
where $\Theta(q)$ is the two-particle scattering phase shift (see below). The function $\sigma(k)$ in the second equation in~\eq{BA 1} is the derivative of the energy function introduced in Ref.~[\onlinecite{Yang}]. It obeys the Yang-Yang equation\cite{Yang,KBI,Sutherland_book}
\beq
\sigma(q) + \frac{1}{2\pi}\!\int_{-q_0}^{q_0}\!dq' \Theta'(q - q')\sigma(q') 
= \frac{\hbar^2 q}{m}.
\label{BA 4} 
\eeq

The functions $\rho(q)$ and $\sigma(q)$ are, respectively, even and odd functions of their argument.\cite{KBI,Sutherland_book} Their values at the Fermi rapidity $\rho_0 = \rho(q_0)$ and $\sigma_0 = \sigma(q_0)$ satisfy\cite{KBI,Sutherland_book}
\beq
\frac{\sigma_0}{\rho_0} = 2\pi\hbar v,
\quad
\sigma_0\rho_0 = \frac{\,\hbar^2 n_0}{2m\,}.
\label{BA 5}
\eeq
Taking into account the relation between the sound velocity $v$ and the Luttinger liquid parameter $K$ [see \Eq{LL 9}], we obtain
\beq
\rho_0 = \frac{\sqrt{K\,}}{2\pi}\,,
\quad
\sigma_0 = \hbar v\sqrt{K\,}.
\label{BA 6}
\eeq

The two-particle scattering phase shift $\Theta(q)$ in Eqs.~\eq{BA 3}  and \eq{BA 4}  is given by\cite{Lieb,KBI,Sutherland_book}
\beq
\Theta(q) = - \,2\arctan(q/c)
\label{BA 7}
\eeq
for the Lieb-Liniger model \eq{LL 1}, and by\cite{Sutherland,Sutherland_book}
\beq
\Theta(q) = 2_{}\text{Im}\!\left[\ln\Gamma\!\left(\lambda + \frac{ia_0 q}{2}\right) 
- \ln\Gamma\!\left(1 + \frac{ia_0 q}{2}\right)\right]
\label{BA 8}
\eeq
for the hyperbolic Calogero-Sutherland model \eq{T 5}. At $|q|\ll\lambda/a_0$ (for $\alpha$ and $\lambda$ satisfying Eqs.~\eq{T 4} and \eq{T 8}, this range includes $|q|\sim q_0$), \Eq{BA 8} simplifies to 
\beq
\Theta(q) = a_0 q \ln\lambda  
- 2_{}\text{Im}\ln\Gamma\!\left(1 + \frac{ia_0 q}{2}\right).
\label{BA 9}
\eeq
The approximation \eq{BA 9} is adequate for excitations with not too high energies, such as $\varepsilon\sim vp_\ast$, and it is equivalent\cite{Sutherland,Sutherland_book,Sklyanin} to neglecting the difference between the hyperbolic Calogero-Sutherland model and the quantum Toda model. For brevity, we shall refer to the results obtained in the framework of the approximation~\eq{BA 9} as pertaining to the quantum Toda model.

The dependence on $q$ in both Eqs.~\eq{BA 7} and \eq{BA 9} is characterized by a single scale $q_*$, with $q_* = c$ for the Lieb-Liniger model and $q_* = 2/a_0$ for the hyperbolic Calogero-Sutherland model. In the regimes we consider, these scales are small compared with the respective values of the Fermi rapidities,
\beq
\frac{q_\ast}{q_0} \sim \zeta\ll 1, 
\label{BA 10}
\eeq
where $\zeta$ is the small parameter introduced in \Sec{models}, see Eqs.~\eq{M 1}, \eq{LL 21}, and \eq{T 19}.
Moreover, it follows from Eqs.~\eq{LL 9}, \eq{LL 19}, \eq{T 15}, \eq{T 18}, and \eq{BA 6} that 
\beq
p_\ast = 6\hbar\rho_0 q_\ast,
\label{BA 11}
\eeq
for both models considered. Equations \eq{BA 1} then show that \mbox{$|q-q_0| \sim q_\ast$} corresponds to $p(q)\sim p_\ast$ and \mbox{$\varepsilon(q) \sim vp_\ast$}. Thus, in order to study the excitation spectra at the classical-to-quantum crossover, it is sufficient to find $\rho$ and $\sigma$ in the vicinity of one of the Fermi rapidities, say, at $q\approx q_0$. Accordingly, it is convenient to work with the ``shifted'' dimensionless rapidities 
\beq
t = \frac{q - q_0}{q_\ast}
\label{BA 12}
\eeq
instead of $q$. 

In terms of $t$, the classical-to-quantum crossover \mbox{at $p\sim p_\ast$} corresponds to $|t|\sim 1$, which is well within the range $|t|\ll q_0/q_\ast \sim 1/\zeta$. At such $t$, the difference between the normalized functions 
\beq
\varrho(t) = \frac{\rho(t)}{\rho_0},
\quad
\varsigma(t) = \frac{\sigma(t)}{\sigma_0}
\label{BA 13}
\eeq
is very small, of order $\zeta\ll 1$. However, this difference cannot be neglected as it is responsible for the nonlinear corrections to the excitation spectra. Indeed, with the help of Eqs.~\eq{BA 11}-\eq{BA 13}, Eqs.~\eq{BA 1} can be cast in the form
\begin{subequations}
\label{BA 14}
\beqa
\frac{p\,}{\,p_\ast} &=& \pm\frac{\pi}{3}\!\int_0^{\pm\tau}\!dt\,\varrho(t),
\label{BA 14a} 
\\
\frac{\varepsilon_\pm - vp\,}{\,vp_\ast} &=& \pm\frac{\pi}{3}\!\int_0^{\pm\tau}\!dt
\bigl[\varsigma(t) - \varrho(t)\bigr],
\label{BA 14b} 
\eeqa
\end{subequations}
where $0 < \tau \ll 1/\zeta$ and the $+ \,(-)$ signs correspond to the type I (type II) excitation. 

\subsection{Crossover functions and their properties}

According to the discussion in Secs.~\ref{KdV} and \ref{models}, the excitation spectra at $\zeta\to 0$ are given by \Eq{M 2}, which we write here as 
\beq
\varepsilon_\pm(p) = vp_\ast\!\left[s + \zeta e_\pm(s) + O\bigl(\zeta^2\bigr)\right], 
\quad
s = p/p_\ast.
\label{BA 15}
\eeq
Comparison of \Eq{BA 15} with Eqs.~\eq{BA 14} then yields the dimensionless crossover functions $e_\pm(s)$ in the form 
\beq
e_\pm(s) = \pm \,E(\pm \,s),
\label{BA 16}
\eeq
where $s > 0$ and the function $E(S)$ is defined parametrically by
\beq
S(\tau) = \frac{\pi}{3}\!\int_0^{\tau}\!\!dt\,\varrho_0(t),
\quad
E(\tau) = \frac{\pi}{3}\!\int_0^{\tau}\!\!dt\,\eta_0(t).
\label{BA 17}
\eeq
Here $\tau$ may have either sign, and the functions $\varrho_0(t)$ and $\eta_0(t)$ are given by
\begin{subequations}
\label{BA 18}
\beqa
\varrho_0(t) &=& \lim_{\zeta\to \,0}\varrho(t) = \lim_{\zeta\to \,0}\varsigma(t),
\label{BA 18a}
\\
\eta_0(t) &=& \lim_{\zeta\to \,0}\frac{1}{\zeta}\bigl[\varsigma(t) - \varrho(t)\bigr],
\label{BA 18b}
\eeqa
\end{subequations} 
with the limits $\zeta\to 0$ evaluated at fixed $t$. 

Further insight is provided by the asymptotic solutions of the Bethe ansatz equations \eq{BA 3}  and \eq{BA 4} at $q$ satisfying \mbox{$q_\ast\ll |q - q_0|\ll q_0$}, or, equivalently, at $t$ in the range $1\ll |t |\ll 1/\zeta$, see Eqs.~\eq{BA 10} and \eq{BA 12}.  As discussed above, this range corresponds to the classical regime in the excitations spectra. Such classical solutions have been found in Refs.~[\onlinecite{Lieb,KMF}] for the Lieb-Liniger model and in Ref.~[\onlinecite{Sutherland}] (see also Ref.~[\onlinecite{Sutherland_book}] for a review) for the quantum Toda model. These solutions and the approximations involved are reviewed in Appendix~\ref{Appendix_Classical}. Substituting the classical solutions into Eqs.~\eq{BA 18}, we obtain
\beq
\varrho_0(t) = \left\lbrace\begin{array}{cr}
(\pi t)^{-1/2}, & ~t\gg 1, 
\\
\bigl|4t/\pi\bigr|^{1/2}, & ~-\,t \gg 1\;
\end{array}\right.
\label{BA 19}
\eeq
for the Lieb-Liniger model and
\beq
\varrho_0(t) = \left\lbrace\begin{array}{cr}
\bigl(4t/\pi\bigr)^{1/2}, & t \gg 1,
\\
|\pi t|^{-1/2}, & -\,t\gg 1\;
\end{array}\right.
\label{BA 20} 
\eeq
for the quantum Toda model, and the relation
\beq
\eta_0(t) = \frac{2\pi}{3}\!\int_0^t\!dt'\varrho_0(t') = 2S(t),
\label{BA 21}
\eeq
valid for both models. The relation \eq{BA 21} turns out to be applicable at \mbox{all $t$}, not only at \mbox{$|t|\gg 1$}. This can be shown rigorously, see Appendix~\ref{Appendix_Beyond}. 

With \Eq{BA 21} at hand, the task of evaluating the crossover functions reduces to finding the function $\varrho_0(t)$ introduced in \Eq{BA 18a}. We therefore turn to the Lieb equation~\eq{BA 3}, divide it by $\rho_0$ [see Eqs.~\eq{BA 6} and \eq{BA 13}], change the variable according to \Eq{BA 12}, and write the resulting equation as
\beq
\varrho(t) + \frac{1}{2\pi}\!\int_{-2k_0}^{0}\!\!dt'\Theta'(t - t')\varrho(t') = \frac{1}{2\pi\rho_0},
\label{BA 22}
\eeq
where $k_0 = q_0/q_\ast\sim 1/\zeta$, see \Eq{BA 10}. 

For the Lieb-Liniger model, the kernel $\Theta'(t)$ in \Eq{BA 22} is given by
\beq
\Theta'(t) = -\,\frac{2\,}{1+ t^2}\,,
\label{BA 23}
\eeq 
whereas $1/\rho_0\propto \zeta^{1/2}$. Accordingly, in the limit $\zeta\to 0$ \Eq{BA 22} turns into a homogeneous equation,
\beq
\varrho_0(t) + \frac{1}{2\pi}\!\int_{-\,\infty}^{0}\!\!dt'\Theta'(t - t')\varrho_0(t') = 0.
\label{BA 24}
\eeq
Augmented with the condition $\varrho_0(0) = 1$, \Eq{BA 24} has a unique solution for the normalized density $\varrho_0(t)$. This solution is derived in Appendix~\ref{Appendix_Wiener-Hopf}. At large $|t|$ it agrees with \Eq{BA 19}, as expected. Note that 
\mbox{$\Theta'(t - t') \varrho_0(t')\bigr|_{t'\to-\,\infty}\propto |t'|^{-3/2}$}, hence the integral on the left-hand side of \Eq{BA 24} converges.

For the quantum Toda model 
\beq
\Theta'(t) = 2\ln\lambda - 2_{}\text{Re}\,\psi\!\left(1 + it\right),
\label{BA 25}
\eeq 
where $\psi(z) = d\ln\Gamma(z)/dz$ is the digamma function.  
With \Eq{BA 20} taken into account, this gives \mbox{$\Theta'(t - t')\varrho_0(t')\bigr|_{t'\to-\,\infty}\propto |t'|^{-1/2}\ln |\lambda/t'|$}, and the integral in the analog of \Eq{BA 24} would be divergent. 
This difficulty is circumvented by first differentiating \Eq{BA 22} with respect to $t$, and only then taking the limit $\zeta\to 0$. The resulting integro-differential equation
\beq
\varrho_0^{\,\prime}(t) 
+ \frac{1}{2\pi}\!\int_{-\infty}^{0}\!\!dt'\Theta^{\prime\prime}(t - t')\varrho_0(t') = 0
\label{BA 26} 
\eeq
and the condition  $\varrho_0(0) = 1$ define $\varrho_0(t)$ uniquely. 

\begin{figure}[t]
\centering
\includegraphics[width=.99\columnwidth]{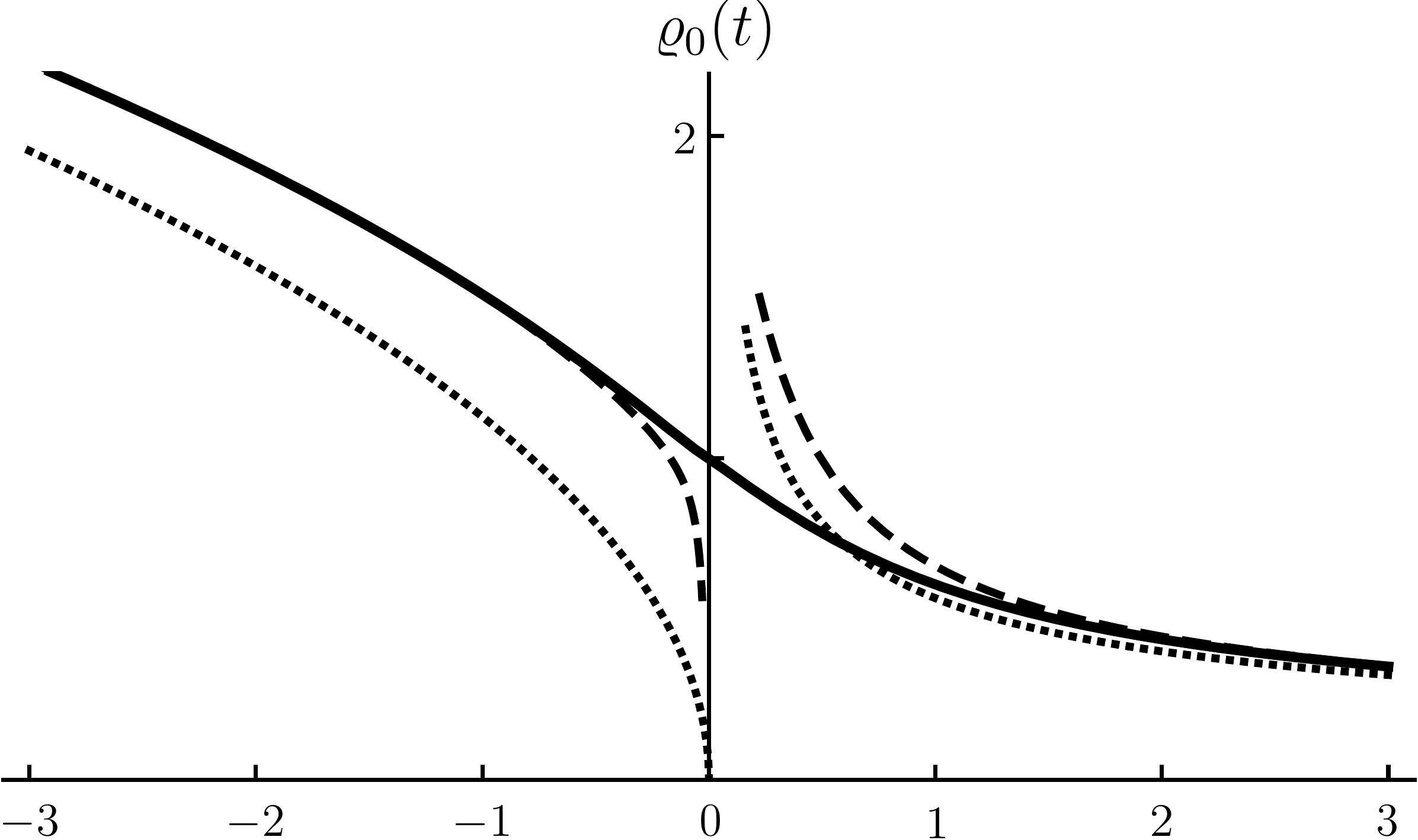}
\caption{
The function $\varrho_0(t)$ for the Lieb-Liniger model. The solid line is a plot of the exact result given by Eqs.~\eq{BA 30}. The dotted lines represent \Eq{BA 19}, applicable at \mbox{$|t|\gg 1$}, i.e., in the classical regime. The dashed lines correspond to~Eqs.~\eq{BA 31}, which include the leading quantum corrections.
}
\label{fig_rho}
\end{figure}

Although the kernels \eq{BA 23} and \eq{BA 25} have quite different appearance, their Fourier transforms turn out to be closely related, as are the functions $\varrho_0(t)$ defined by Eqs.~\eq{BA 24} and \eq{BA 26}. Indeed, as shown in Appendix~\ref{LL vs QT},
\beq
\varrho_0(t)\bigr|_\text{Lieb-Liniger} = \varrho_0(-t)\bigr|_\text{quantum Toda}.
\label{BA 27}
\eeq 
In view of Eqs.~\eq{BA 17} and \eq{BA 21}, \Eq{BA 27} translates to the relation
\beq
E(S)\bigr|_\text{Lieb-Liniger} 
= E(-\,S)\bigr|_\text{quantum Toda},
\label{BA 28}
\eeq
and \Eq{BA 16} yields
\beq
e_\pm (s)\bigr|_\text{Lieb-Liniger} = -\,e_\mp(s)\bigr|_\text{quantum Toda},
\label{BA 29}
\eeq
in agreement with the results of Secs.~\ref{KdV} and \ref{models}.

\subsection{Evaluation of the crossover functions}
\label{evaluation}

Equation~\eq{BA 24} is of Wiener-Hopf type and can be solved analytically, see Appendix~\ref{Appendix_Wiener-Hopf}. The solution satisfying $\varrho_0(0) = 1$ can be written as
\begin{subequations}
\label{BA 30}
\beq
\varrho_0(t) = \frac{1}{\pi\sqrt{2\pi}}\!
\int_0^\infty\!\!\frac{dz}{z^{1/2}}
\sin(2\pi z)\Gamma(z)\,e^{-z(\ln z -1+ 2\pi t)} 
\label{BA 30a}
\eeq
at $t > 0$ and 
\beq
\varrho_0(t) = \frac{1}{\pi\sqrt{2\pi}}
\fint_0^\infty\!\!\!\frac{dz}{z^{3/2}}\!
\left[1 - \frac{\pi e^{z(\ln z -1+ 2\pi t)}}{\tan(\pi z)\Gamma(z)}\right]
\label{BA 30b}
\eeq
\end{subequations}
at $t < 0$. Simple poles at integer $z$ in the integrand of \Eq{BA 30b} are understood in the Cauchy principal value sense. The function $\varrho_0(t)$ is plotted in \Fig{fig_rho}.

\begin{figure}[t]
\centering
\includegraphics[width=.99\columnwidth]{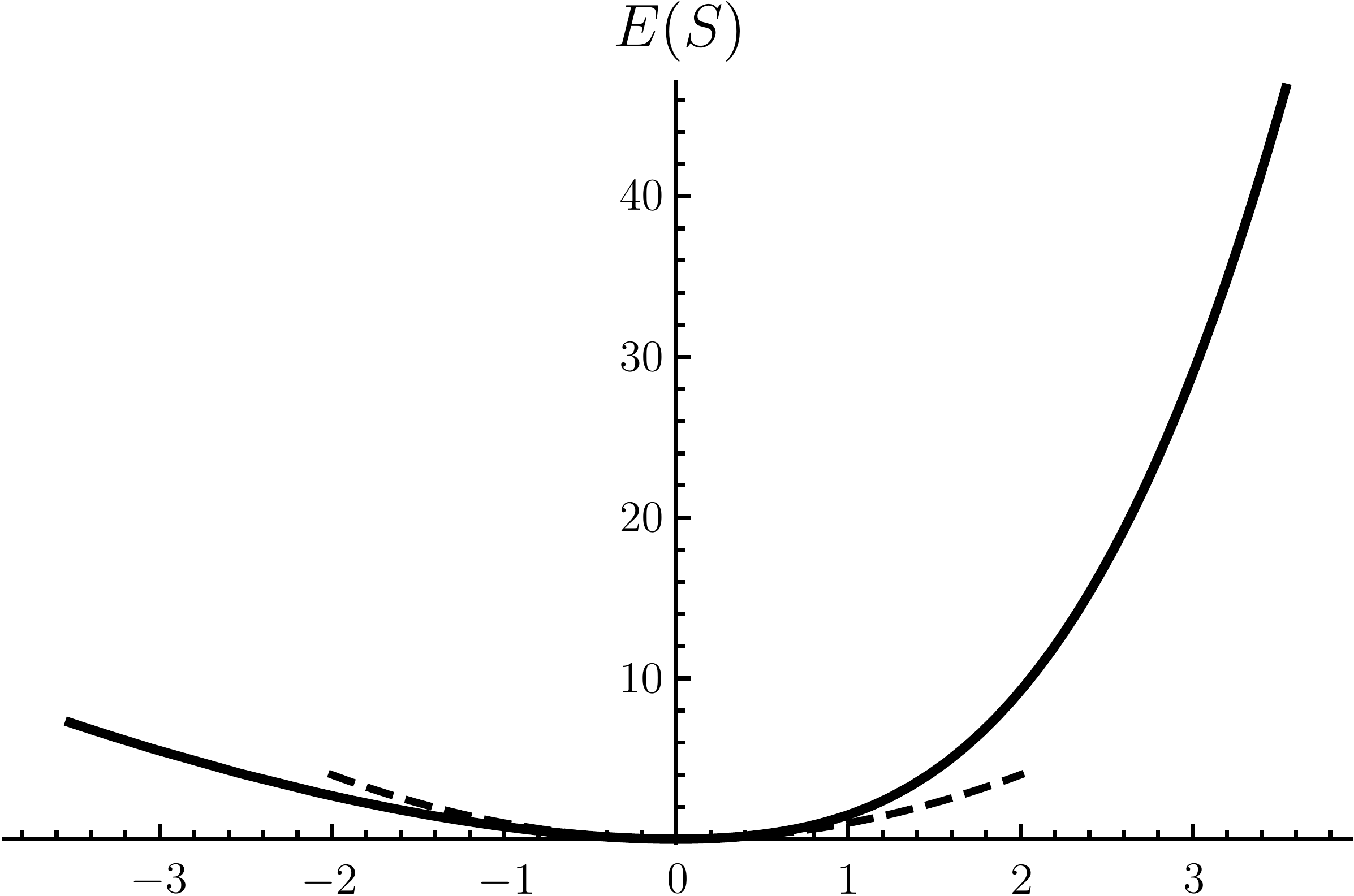}
\caption{
The function $E(S)$ for the Lieb-Liniger model. The solid line is a plot of the exact result obtained by substituting Eqs.~\eq{BA 30} into Eqs.~\eq{BA 17} and \eq{BA 21}. The dashed line is a plot of \mbox{$E(S) = S^2$}, applicable in the quantum regime $|S|\ll 1$ [see \Eq{BA 36}].
}
\label{fig_E(S)}
\end{figure}

As expected for a solution of an integral equation with a non-singular kernel [see Eqs.~\eq{BA 23} and \eq{BA 24}], $\varrho_0(t)$ is an analytic function. It is easy to check that $\varrho_0(t) > 0$ and \mbox{$d\varrho_0(t)/dt < 0$}. The first of these inequalities implies that the function $E(S)$ defined by Eqs.~\eq{BA 17} and \eq{BA 21} is analytic. This function is plotted in \Fig{fig_E(S)}. It satisfies \mbox{$E^{\prime\prime}(S) > 0$} at all $S$ and \mbox{$E(0) = E'(0) = 0$}. Accordingly, the crossover functions $e_\pm(s)$ [see \Eq{BA 16}] and their derivatives $e^\prime_\pm(s)$ vanish at $s = 0$, whereas $e^\prime_+(s) > 0$ and $e^\prime_-(s) < 0$ at \mbox{finite $s$}. 

The asymptotes of the crossover functions in the classical ($s\gg 1$) and quantum ($s\ll 1$) regimes can be found analytically. At large $|t|$, Eqs.~\eq{BA 30} yield
\begin{subequations}
\label{BA 31}
\beqa
&&\negquad\varrho_0(t)\bigr|_{t\gg 1} = (\pi t)^{-1/2}\!\left[1+ \dfrac{\ln(8\pi t) -1}{4\pi t} + \ldots\right],
\qquad\quad
\label{BA 31a} \\
&&\negquad\varrho_0(t)\bigr|_{-t\gg 1} = \left|\dfrac{4t}{\pi}\right|^{1/2}\!\left[
1 + \dfrac{\ln|8\pi t| + 1}{|4\pi t|}
+ \ldots\right],
\label{BA 31b}
\eeqa
\end{subequations}
in agreement with \Eq{BA 19}. Substitution into Eqs.~\eq{BA 17} and \eq{BA 21} then gives
\begin{subequations}
\label{BA 32}
\beqa
\!E(S)\bigr|_{S\gg 1} &=&\! S^3 + \frac{2}{3}\,S + \ldots,
\qquad\qquad\qquad\qquad\quad
\label{BA 32a}\\
\!E(S)\bigr|_{-\,S\gg 1} 
\!&=&\! \frac{3}{5}\!\left(\frac{2\pi}{3}\right)^{\!2/3}\!|S|^{5/3} -\frac{2}{9}\,|S| + \ldots.
\label{BA 32b}
\eeqa
\end{subequations}
Interestingly, the logarithmic terms, dominating the corrections to $\varrho_0(t)$ in Eqs.~\eq{BA 31}, do not contribute to the expansions \eq{BA 32}.

Taking into account Eqs.~\eq{BA 16} and \eq{BA 29}, we obtain the asymptotes of the crossover functions in the classical regime \mbox{$s\gg 1$},
\begin{subequations}
\label{BA 33}
\beqa
e_+(s) &=& s^3 + \frac{2}{3}\,s + \ldots,
\qquad\qquad\qquad\qquad
\label{BA 33a} \\
e_-(s) &=& -\,\frac{3}{5}\!\left(\frac{2\pi}{3}\right)^{\!2/3}\!s^{5/3} + \frac{2}{9}\,s + \ldots
\label{BA 33b}
\eeqa
\end{subequations}
for the Lieb-Liniger model, and
\begin{subequations}
\label{BA 34}
\beqa
e_+(s) &=& \frac{3}{5}\!\left(\frac{2\pi}{3}\right)^{\!2/3}\!s^{5/3}  - \frac{2}{9}\,s + \ldots,
\qquad
\label{BA 34a} \\
e_-(s) &=& - s^3 - \frac{2}{3}\,s + \ldots
\label{BA 34b}
\eeqa
\end{subequations}
for the quantum Toda model.  
The first terms on the right-hand sides of Eqs.~\eq{BA 33} and \eq{BA 34} can be deduced from the solutions of the classical equation of motion, see Eqs.~\eq{KdV 20} and \eq{KdV 21}. The second terms in Eqs.~\eq{BA 33} and \eq{BA 34} represent the leading quantum corrections.

At small $|t|$, the function $\varrho_0(t)$ can be expanded in Taylor series. The first two terms of the expansion read
\beq
\varrho_0(t)\bigr|_{|t|\ll 1} = 1 - \pi t/6 + \ldots
\label{BA 35}
\eeq
(see Appendix~\ref{Appendix_Wiener-Hopf}), and Eqs.~\eq{BA 17} and \eq{BA 21} result in
\beq
E(S)\bigr|_{|S|\,\ll \,1} = S^2 + \frac{1}{3}S^3 + \ldots,
\label{BA 36}
\eeq
which yields
\beq
e_\pm(s) = \pm\,s^2 + \frac{1}{3} s^3 +\ldots
\label{BA 37}
\eeq
for the Lieb-Liniger model and
\beq
e_\pm(s) = \pm\,s^2 - \frac{1}{3} s^3 +\ldots
\label{BA 38}
\eeq
for the quantum Toda model in the quantum regime \mbox{$s \ll 1$}. The leading contributions in Eqs.~\eq{BA 37} and~\eq{BA 38} correspond to the particle and hole excitations of a gas of free fermions with a quadratic dispersion relation, see \Eq{KdV 26}. 

\section{Discussion}
\label{Discussion}

The regime in which the interaction between particles dominates the energy of a quantum system is often referred to as classical. Indeed, many properties of a quantum system in this regime can be understood by solving the corresponding classical equations of motion. 
Paradigmatic examples of one-dimensional systems exhibiting such semi-classical behavior are identical bosons with a weak short-range repulsion and identical particles, either bosons or fermions, with a strong long-range repulsion. Despite the obvious difference between these two families of systems, it can be shown that their low-energy excitations admit a universal description in terms of the quantized version of the so-called first Hamiltonian structure\cite{Soliton_books} of the KdV equation. This observation implies, in particular, that the spectra of elementary excitations for the members of the two families are related to that of the quantum KdV model~\eq{KdV 10}, and, therefore, to each other, see \Eq{M 2}. 

The relevant classical equation of motion is the celebrated KdV equation \eq{KdV 1}. This equation has two physically and mathematically distinct types of solutions: the delocalized periodic waves, corresponding to phonons in quantum systems, and the solitons. The classical solutions translate to power-law excitation spectra characterized by different exponents for phonons and solitons, see Eqs.~\eq{KdV 19}-\eq{KdV 21}. Importantly, in quantum systems these two excitation branches are not independent, and their spectra can be viewed as analytical continuations of one another [see Eqs.~\eq{BA 15} and \eq{BA 16}], with the classical expressions \eq{KdV 20} and \eq{KdV 21} serving as high-energy asymptotes. 

At the lowest energies, the classical treatment inevitably breaks down. Instead, the system is best described in terms of weakly interacting fermions with a quadratic spectrum.\cite{Rozhkov,ISG_RMP} These effective fermions can be used to characterize the elementary excitations in the classical regime as well. Indeed, as discussed in \Sec{KdV}, the bright (dark) solitons carry a macroscopic number of fermionic particles (holes). With the decrease of the soliton's energy, this number decreases, becoming unity when the bright (dark) soliton reaches its ultimate quantum fate of being demoted to a single particle (hole) excitation of the effective Fermi gas. 

In this paper we considered in detail the integrable members of the two families of systems mentioned above, namely, the Lieb-Liniger model \eq{LL 1} and the quantum Toda model \eq{T 1}. 
The key advantage of working with integrable models is that their excitation spectra can be found analytically, yielding explicit expressions valid throughout the classical-to-quantum crossover, see \Sec{Excitations}. The spectra we found satisfy Eqs.~\eq{BA 15} and~\eq{BA 29}, as expected for models with quantum KdV-type low-energy behavior (see \Sec{models}). 

Our results can be reformulated in terms of the exact spectra of the elementary excitations of the quantum KdV model defined by Eqs.~\eq{KdV 7}, \eq{KdV 9}, and \eq{KdV 10},
\beq
\varepsilon_\pm(p) = \pm\,\varepsilon_\ast E\bigl(\mp \,p/p_\ast\bigr).
\label{D 1}
\eeq
Here $\varepsilon_\pm$ and $p$ are eigenvalues of the operators $H_\text{KdV}$ [see \Eq{KdV 10}] and $P$ [see \Eq{KdV 9}], and $\varepsilon_\ast$ and $p_\ast$ are the energy and momentum scales expressed via the parameters $m_\ast$ and $a_\ast$ of $H_\text{KdV}$ according to \Eq{KdV 18}.  The dimensionless function $E(S)$ in \Eq{D 1} is defined parametrically by Eqs.~\eq{BA 17}, \eq{BA 21}, and \eq{BA 30}. This function is plotted in \Fig{fig_E(S)}, and its asymptotes are given in Eqs.~\eq{BA 32} and \eq{BA 36}. Because the quantum KdV Hamiltonian~\eq{KdV 10} is chiral, the two excitation branches $\varepsilon_\pm(p)$ coincide with the bounds on the excitation energy~$\varepsilon$ at a given momentum $p$: $\varepsilon_-(p)\leq\varepsilon\leq\varepsilon_+(p)$. 

Finally, we note that the derivation of the low-energy quantum KdV theories for the Lieb-Liniger and the quantum Toda models in \Sec{models} does not rely on their integrability and can be readily adapted for generic (i.e., non-integrable) systems.\cite{PM_WignerToda} In such systems~$\varepsilon_+(p)$ excitation mode no longer represents the exact eigenstate and, therefore, has a finite lifetime even at zero temperature.\cite{ISG_RMP,KPKG_fermions,MF,bosons_decay,PM_WignerToda,RM} 
However, the resulting inelastic broadening is parametrically small near the classical limit.\cite{PM_WignerToda,RM} Therefore, we expect our results for the excitation spectra to be applicable to all quantum one-dimensional systems with a classical limit described by the KdV equation.

\begin{acknowledgments}
We benefited from discussions with A. G. Abanov. 
This work was supported by the US Department of Energy, Office of Science, Materials Sciences and Engineering Division. We are grateful to the Aspen Center for Physics (NSF Grant No. PHYS-1066293) for hospitality.
\end{acknowledgments}

\appendix
\section{Classical regime from Bethe ansatz}
\label{Appendix_Classical}

In this Appendix, we review asymptotic solutions\cite{Lieb,Sutherland,KMF,Sutherland_book} of the Bethe ansatz equations~\eq{BA 3} and \eq{BA 4} at $q_\ast/q_0\sim \zeta$ approaching zero. Here $q_\ast = c$ for the Lieb-Liniger model and $q_\ast = 2/a_0$ for the hyperbolic Calogero-Sutherland model in the Toda limit.

It is convenient to work with dimensionless rapidities \mbox{$k = q/q_\ast$}. After such rescaling, the Lieb equation \eq{BA 3} and the Yang-Yang equation \eq{BA 4} become
\begin{subequations}
\label{A 1}
\beqa
\rho(k) + \frac{1}{2\pi}\!\int_{-k_0}^{k_0}\!\!dk'\Theta'(k - k')\rho(k') &=& \frac{1}{2\pi},
\label{A 1a} \\
\sigma(k) + \frac{1}{2\pi}\!\int_{-k_0}^{k_0}\!dk' \Theta'(k - k')\sigma(k') 
&=& \frac{\hbar^2 q_\ast}{m_\past} k.
\qquad
\label{A 1b} 
\eeqa
\end{subequations}
The dimensionless Fermi rapidity $k_0 = q_0/q_\ast$ is to be determined self-consistently from the normalization condition
\beq
\int_{-k_0}^{k_0}\!\! dk\,\rho(k) = \frac{n_0}{q_\ast},
\label{A 2} 
\eeq
see \Eq{BA 2}. In the regimes we consider $k_0 \sim 1/\zeta\gg 1$. 

We are interested in the behavior of the functions $\rho(k)$ and $\sigma(k)$ at $k$ in the range $\bigl||k| - k_0\bigr|\gg 1$. The momenta $p(k)$ corresponding to such $k$ satisfy \mbox{$p(k)\gg p_\ast$}, i.e., they belong to the classical regime in the excitation spectra. 

\subsection{Lieb-Liniger model}

For $k_0\gg 1$ the integrals on the left-hand sides of Eqs.~\eq{A 1} are dominated by $k'$ satisfying $k_0 - |k'|\gg 1$. Moreover, it is natural to assume that both at $k_0 - |k|\gg 1$ and at $|k| - k_0\gg 1$ the functions $\rho(k)$ and $\sigma(k)$ vary with $k$ on the scale of order $k_0$ as no other scale is available. Therefore, in order to find $\rho(k)$ and $\sigma(k)$ at $\bigl||k| - k_0\bigr|\gg 1$, it is sufficient to replace the two-particle scattering phase shift \mbox{$\Theta(k) = -\,2\arctan k$} [see \Eq{BA 7}] in Eqs.~\eq{A 1} by its asymptote 
\beq
\widetilde\Theta(k) = \Theta(k)\bigr|_{|k|\gg 1} = -\,\pi\,\sgn(k) + \frac{2}{k}\,.
\label{A LL 1} 
\eeq
With this approximation, \Eq{A 1a} simplifies to
\beq
\theta\bigl(|k| - k_0\bigr)\rho(k) 
+ \frac{1}{\pi}\frac{d}{dk}\fint_{-k_0}^{k_0}\!\!dk'\frac{\,\rho(k')}{k - k'} = \frac{1}{2\pi}\,,
\label{A LL 2} 
\eeq
with the pole in the integrand treated as the Cauchy principal value. 

Unlike the exact phase shift $\Theta(k)$, the approximate phase shift $\widetilde\Theta(k)$ has a singularity at $k = 0$. This singularity translates to non-analyticities at $k\to\pm\,k_0$ in the solutions of the approximate Lieb equation \eq{A LL 2}. In fact, the interior of the interval \mbox{$|k| < k_0$} does not enter \Eq{A LL 2} on an equal footing with its exterior: \Eq{A LL 2} can be viewed both as an equation for $\rho(k)$ in the interior, and as a prescription for extending the solution from the interior to the exterior.  

Another difficulty that stems from the singular behavior of $\widetilde\Theta(k)$ is that \Eq{A LL 2} does not determine $\rho(k)$ at $|k| < k_0$ uniquely: if $\rho(k)$ is a solution, then $\rho + \delta\rho$ with $\delta\rho\propto (k_0^2 - k^2)^{-1/2}$ is a solution as well. The ambiguity can be removed by imposing appropriate boundary conditions. For $\zeta\to 0$, such conditions read\cite{Lieb}
\beq
\rho(k)\bigr|_{k \,=\, \pm (k_0 - 0)} = 0.
\label{A LL 3}  
\eeq
The conditions~\eq{A LL 3} are justified by the observation that the exact density $\rho(k)$ near the boundaries of the interval $|k| < k_0$ is parametrically smaller than that in its interior. Indeed, with the help of \Eq{A LL 2}, $\rho(k)$ at $k_0 - |k| \gg 1$ is estimated as $\rho(k) \sim \rho(0) \sim k_0$. Substituting this estimate into \Eq{A 2}, we find \mbox{$k_0 \sim\gamma^{-1/2}$}. On the other hand, at \mbox{$k_0 - |k| \lesssim 1$} [recall that such $k$ are not handled properly by the approximate equation \eq{A LL 2}] we have $\rho(k) \sim\rho_0$, where \mbox{$\rho_0\sim K^{1/2}\!\sim\gamma^{-1/4}$} is the exact value of $\rho(k)$ at the Fermi rapidity, see Eqs.~\eq{BA 6} and \eq{LL 9}. Thus, the ratio of $\rho(k)$ near the boundaries to that in the interior is indeed small, of order $\gamma^{1/4}\!\sim\zeta^{1/2}$ [see Eqs.~\eq{LL 9} and~\eq{LL 21}]. 
 
The solution of \Eq{A LL 2} subject to the conditions \eq{A LL 3} is unique and reads\cite{Lieb,KMF}
\beq
\rho(k) = \dfrac{1}{2\pi} \!\times\! \left\lbrace
\begin{array}{cr} 
\bigl(k_0^2 - k^2\bigr)^{1/2},\qquad
& |k| < k_0,
\\
\\
\sgn(k) 
\dfrac{d}{dk} \bigl(k^2 - k_0^2\bigr)^{1/2},
& |k| > k_0.
\end{array}
\right.
\label{A LL 4} 
\eeq
Substituting \Eq{A LL 4}  into \Eq{A 2} and taking into account  Eqs.~\eq{LL 9} and \eq{LL 21}, we obtain
\beq
k_0 = \frac{\,2}{\sqrt{\gamma\,}} = \frac{2K}{\pi} = \frac{9}{4\pi\zeta}
\label{A LL 5} 
\eeq
for the dimensionless Fermi rapidity, in agreement with the above estimate. 

We now focus on the realm of the long-wavelength excitations, where the universal KdV description is expected to be applicable. For such excitations $k$ is close to $k_0$, and \Eq{A LL 4} can be expanded in powers of \mbox{$|k - k_0|/k_0$}, leading to
\begin{subequations}
\label{A LL 6}
\beq
\rho(k) = \frac{k_0}{\pi}\!
\left[\left(\frac{k_0 - k}{\,2k_0}\right)^{1/2} \!- \frac{1}{2}\!\left(\frac{k_0 - k}{\,2k_0}\right)^{3/2} \!+ \ldots\,\right]
\qquad~
\label{A LL 6a}
\eeq
at $k < k_0$ and
\beqa
\rho(k) = \frac{1}{4\pi}\!
\left[\left(\frac{k - k_0}{\,2k_0}\right)^{\!-1/2} \!\!+ \frac{3}{2}\!\left(\frac{k - k_0}{\,2k_0}\right)^{1/2} \!\!+ \ldots\,\right]
\quad~
\label{A LL 6b}
\eeqa
\end{subequations}
at $k > k_0$. Changing the variable to \mbox{$t = k - k_0$} [see \Eq{BA 12}] and taking into account Eqs.~\eq{BA 6}, \eq{BA 13}, and \eq{A LL 5}, we rewrite Eqs.~\eq{A LL 6} in a more compact form as 
\beq
\varrho(t) = \frac{\rho(t)}{\rho_0} 
= \varrho_0(t) + \frac{\pi\zeta}{6}\!\int_0^t\!dt'\varrho_0(t')\,+\,\ldots,
\label{A LL 7}
\eeq
where
\beq
\varrho_0(t) = \theta(-t) \bigl|4t/\pi\bigr|^{1/2} \!+ \theta(t)(\pi t)^{-1/2}.
\label{A LL 8}
\eeq

Equation \eq{A LL 7} represents the expansion in small \mbox{$\zeta t\sim (k-k_0)/k_0$} of the dominant at $\zeta\to 0$ and fixed $\zeta t$ contribution to the normalized density $\varrho$. Alternatively, \Eq{A LL 7} can be interpreted as an asymptotic expansion in small~$\zeta$ of $\varrho(t)$ evaluated at fixed $t$, with $t$-dependent expansion coefficients found at $|t|\gg 1$, cf.~Ref~[\onlinecite{Popov_LL}]. Although in such an interpretation Eqs.~\eq{A LL 7} and \eq{A LL 8} are applicable only at large $|t|$, it is reassuring that at $|t|\sim 1$ they yield the correct estimate $\varrho_0(t)\sim 1$. 

Analysis of the Yang-Yang equation \eq{A 1b} follows the same steps. Approximating $\Theta(k)$ in \Eq{A 1b} by $\widetilde\Theta(k)$ [see \Eq{A LL 1}] and taking into account \Eq{A LL 5}, we obtain the equation
\beq
\theta\bigl(|k| - k_0\bigr)\sigma(k) 
+ \frac{1}{\pi}\frac{d}{dk}\fint_{-k_0}^{k_0}\!\!dk'\frac{\,\sigma(k')}{k - k'} = \frac{2\pi\hbar v k}{k_0}.
\label{A LL 9} 
\eeq
As with \Eq{A LL 2}, \Eq{A LL 9} does not define $\sigma(k)$ uniquely and must be augmented with appropriate boundary conditions at $k \to\pm\,k_0$. At $k_0 - |k|\gg 1$ we have $|\sigma(k)|\sim\hbar v k_0\sim\hbar v \gamma^{-1/2}$, which for $\gamma\sim\zeta^2\ll 1$ is parametrically larger than $\sigma(k)$ at $k$ approaching $k_0$, $\sigma(k)\sim\sigma_0\sim\hbar v\gamma^{-1/4}$, see Eqs.~\eq{BA 6} and \eq{A LL 5}. This observation leads to the condition
\beq
\sigma(k)\bigr|_{k \,=\, \pm (k_0 - 0)} = 0,
\label{A LL 10}
\eeq
cf.~\Eq{A LL 3}. The solution of \Eq{A LL 9} satisfying the conditions \eq{A LL 10} is given by\cite{KMF}
\beq
\hspace{-4.6pt}
\sigma(k) = \dfrac{\hbar v\,}{3k_0}\!\times\! \left\lbrace
\begin{array}{cr}
-\,\dfrac{d}{dk} \bigl(k_0^2 - k^2\bigr)^{3/2}, \qquad
& |k| < k_0,
\\
\\
\sgn(k)\dfrac{d^2}{dk^2} \bigl(k^2 - k_0^2\bigr)^{3/2},
& |k| > k_0.
\end{array}
\right.
\label{A LL 11} 
\eeq
Expanding \Eq{A LL 11} in $(k - k_0)/k_0\sim \zeta t$ and taking into account Eqs.~\eq{BA 6} and \eq{BA 13}, we find
\beq
\varsigma(t) = \frac{\sigma(t)}{\sigma_0}
= \varrho_0(t) + \frac{5\pi\zeta}{6}\!\int_0^t\!dt'\varrho_0(t')\,+\,\ldots
\label{A LL 12}
\eeq
with $\varrho_0(t)$ given by \Eq{A LL 8}. Similar to \Eq{A LL 7}, \Eq{A LL 12} can be viewed as an asymptotic expansion in small $\zeta$ of the function $\varsigma(t)$ evaluated at fixed $t$ such that $|t|\gg 1$, which corresponds to the classical regime in the excitation spectra. Substitution of Eqs.~\eq{A LL 7} and \eq{A LL 12} into \Eq{BA 18b} yields \Eq{BA 21}.

\subsection{Quantum Toda model}

Instead of attacking the quantum Toda model directly, we consider the hyperbolic Calogero-Sutherland model \eq{T 5} in the Toda limit defined by Eqs.~\eq{T 4} and \eq{T 8}. As shown in Ref.~[\onlinecite{Sutherland_book,Sutherland}] and confirmed below, under these conditions the dimensionless Fermi rapidity~$k_0$ in the Bethe ansatz equations \eq{A 1} satisfies
\beq
1\ll k_0\ll\lambda.
\label{A CS 1}
\eeq
Accordingly, for $|k|\sim k_0\ll\lambda$ the two-particle phase shift 
\beq
\Theta(k)  
= 2_{}\text{Im}\bigl[\ln\Gamma\!\left(\lambda + ik\right) - \ln\Gamma\!\left(1 + ik\right)\bigr]
\label{A CS 2}
\eeq
[see \Eq{BA 8}] can be replaced with
\beq
\Theta(k)\bigr|_{|k|\ll\lambda}  
= 2k \ln\lambda - 2_{}\text{Im}\ln\Gamma\!\left(1 + ik\right).
\label{A CS 3}
\eeq
This approximation ignores the existence of the regime $|k|\gg\lambda$, absent in the true quantum Toda model,\cite{Sklyanin} but is adequate for our purpose. On the other hand, similar to the Lieb-Liniger model, the dominant contribution to the integrals on the left-hand sides of Eqs.~\eq{A 1} comes from $k'$ satisfying $k_0 - |k'|\gg 1$. Thus, in order to find $\rho(k)$ and $\sigma(k)$ at $k$ in the range $\bigl||k| - k_0\bigr|\gg 1$, the phase shift can be further approximated by
\beq
\widetilde\Theta(k) = \Theta(k)\bigr|_{1\ll |k|\ll\lambda} 
= 2k\bigl(\ln\!|\lambda/k| + 1\bigr). 
\label{A CS 4} 
\eeq 
The Lieb equation~\eq{A 1a} then assumes the form\cite{Sutherland_book,Sutherland}
\beq
\rho(k) + \frac{1}{\pi}\!\int_{-k_0}^{k_0}\!\!dk'\ln\!\left|\frac{\lambda}{k - k'}\right|\rho(k')
= \frac{1}{2\pi}.
\label{A CS 5} 
\eeq

Similar to \Eq{A LL 2} for the Lieb-Liniger model, \Eq{A CS 5} serves simultaneously as an equation for $\rho(k)$ in the interior of the interval $|k| < k_0$, and as a prescription for extending $\rho(k)$ from the interior to the exterior. Moreover, at $|k|< k_0$ the first term on the left-hand side of \Eq{A CS 5} can be neglected.\cite{Sutherland,Sutherland_book} Indeed, at $k_0 - |k|\gg 1$ this term is of order \mbox{$\rho(k)\sim\rho(0)$}, whereas the second term is parametrically larger, of order $k_0\ln(\lambda/k_0)\rho(0)\gg\rho(0)$. 

With these approximations the Lieb equation can be solved exactly.\cite{Sutherland,Sutherland_book} The solution reads
\beq
\mkern-10mu
\rho(k) = \dfrac{1}{2\pi\ln(2\lambda/k_0)}
\times\!\left\lbrace\!
\begin{array}{cr}
\bigl(k_0^2 - k^2\bigr)^{-1/2},
& |k| < k_0,
\\
\\
\arccosh(k/k_0)\,,
& |k| > k_0.
\end{array}
\right.
\label{A CS 6} 
\eeq
The normalization condition \eq{A 2} and Eqs.~\eq{T 15} and~\eq{T 19} then yield
\beq
k_0 = 2\lambda e^{-\alpha} = \frac{\pi}{2\alpha^2 K} = \frac{3}{4\pi\zeta}
\label{A CS 7} 
\eeq
for the dimensionless Fermi rapidity. Since \mbox{$\lambda \gg e^\alpha\gg 1$} [see Eqs.~\eq{T 4} and \eq{T 8}], $k_0$ indeed satisfies the inequalities \eq{A CS 1}.
In the same approximation, \mbox{$\Theta(k)\approx\widetilde\Theta(k)$} [see \Eq{A CS 4}], the Yang-Yang equation~\eq{A 1b} yields\cite{Sutherland,Sutherland_book}
\beq
\sigma(k) = \frac{\hbar v}{\,\alpha k_0}\!\times\! \left\lbrace
\begin{array}{cr}
-\,\dfrac{d}{dk} \bigl(k_0^2 - k^2\bigr)^{1/2},~
& |k| < k_0,
\\
\\
\sgn(k)\bigl(k^2 - k_0^2\bigr)^{1/2},
& |k| > k_0.
\end{array}
\right.
\label{A CS 8} 
\eeq

Focusing on the long-wavelength excitations, we expand Eqs.~\eq{A CS 6}  and \eq{A CS 8} in powers of small \mbox{$(k - k_0)/k_0 = 4\pi \zeta t/3$}. With Eqs.~\eq{BA 6} and \eq{BA 13} taken into account, the first two terms of these expansions can be written as 
\begin{subequations}
\label{A CS 9}
\beqa
\varrho(t)
&=& \varrho_0(t) - \frac{\pi\zeta}{6}\!\int_0^t\!dt'\varrho_0(t') \,+\,\ldots,
\qquad
\label{A CS 9a} \\
\varsigma(t) 
&=&\varrho_0(t) + \frac{\pi\zeta}{2}\!\int_0^t\!dt'\varrho_0(t') \,+\,\ldots
\label{A CS 9b}
\eeqa
\end{subequations}
with $\varrho_0(t)$ given by 
\beq
\varrho_0(t) = \theta(-t) |\pi t|^{-1/2} + \,\theta(t)\bigl(4t/\pi\bigr)^{1/2}.
\label{A CS 10} 
\eeq
Similar to their Lieb-Liniger model counterparts \eq{A LL 7} and \eq{A LL 12}, Eqs.~\eq{A CS 9} can be interpreted as asymptotic expansions in small $\zeta$ of the functions $\varrho(t)$ and $\varsigma(t)$ evaluated at fixed large $t$. Although the numerical coefficients in front of the second terms on the right-hand sides of Eqs.~\eq{A CS 9} differ from those in Eqs.~\eq{A LL 7} and \eq{A LL 12}, the difference does not affect the form of the function $\eta_0(t)$ [see \Eq{BA 18b}], which is again given by \Eq{BA 21}.

\section{Beyond the classical regime}
\label{Appendix_Beyond}
 
Equation \eq{BA 21} can be obtained by substituting $\varrho(t)$ and $\varsigma(t)$ in the form of Eqs.~\eq{A LL 7} and~\eq{A LL 12} for the Lieb-Liniger model and Eqs.~\eq{A CS 9} for the quantum Toda model into \Eq{BA 18b}. However, such a derivation is valid only at $|t|\gg 1$, i.e., in the classical regime in the excitation spectra, whereas we are interested in $|t|\sim 1$, which corresponds to the classical-to-quantum crossover. It turns out that relaxing the restriction $|t|\gg 1$ requires merely a replacement of the functions $\varrho_0(t)$ in Eqs.~\eq{A LL 7}, \eq{A LL 12},  \eq{A CS 9}, and \eq{BA 21} with the exact solutions of Eqs.~\eq{BA 24} or \eq{BA 26}. In this Appendix, we derive such a generalization of the expansion \eq{A LL 7} for the Lieb-Liniger model. Generalizations of Eqs.~\eq{A LL 12} and \eq{A CS 9} can be obtained in a similar manner. 

It is convenient to write the Lieb equation \eq{BA 22} as
\beq
\mathcal L_{2k_0}[\varrho] = \frac{1}{2\pi\rho_0}.
\label{pt 1}
\eeq
Here $k_0\sim 1/\zeta$, $\rho_0\sim\zeta^{-1/2}$ [see Eqs.~\eq{BA 6} and \eq{A LL 5}], and the functional $\mathcal L_\tau$ is defined by
\beq
\mathcal L_\tau\bigl[f\bigr] = f(t) + \frac{1}{2\pi}\!\int_{-\tau}^{0}\!\!dt'\Theta'(t - t')f(t'),
\label{pt 2}
\eeq
where $\Theta(t) = -\,2\arctan t$. The solution of \Eq{pt 1} can be viewed as a function of two variables, $t$ and $\zeta$. We are interested in the behavior of $\varrho(t,\zeta)$ at arbitrary $t$ and small $\zeta\ll \min\{1, |t|^{-1}\}$.

Replacement of the phase shift $\Theta(t)$ in \Eq{pt 2} with $\widetilde\Theta(t)$ given by \Eq{A LL 1} leads to the approximate Lieb equation 
\beq
\widetilde{\mathcal L}_{2k_0}[\widetilde\varrho\,] = \frac{1}{2\pi\rho_0},
\label{pt 3}
\eeq
which, unlike \Eq{pt 1}, is exactly solvable. To avoid confusion, in this Appendix we use a \textit{tilde} to distinguish the exact solution $\widetilde\varrho(t,\zeta)$ of the approximate Lieb equation~\eq{pt 3} from the solution $\rho(t,\zeta)$ of the exact Lieb equation~\eq{pt 1}. 
At $\zeta |t|\ll 1$, the function $\widetilde\varrho(\zeta,t)$ can be expanded as
\beq
\widetilde\varrho(t,\zeta) 
= \widetilde\varrho_0(t) + \frac{\pi\zeta}{6}\widetilde\varrho_1(t) \,+\,\ldots
\label{pt 4}
\eeq
with $\widetilde\varrho_1(t) = \int_0^t\!dt'\widetilde\varrho_0(t')$ [see \Eq{A LL 7}].
The function $\widetilde\varrho_0(t)$ in \Eq{pt 4} satisfies the equation
\beq
\widetilde{\mathcal L}_{\infty}[\widetilde\varrho_0] = 0
\label{pt 5}
\eeq
and is given by $\widetilde\varrho_0(t) = \theta(-t) \bigl|4t/\pi\bigr|^{1/2} \!+ \theta(t)(\pi t)^{-1/2}$ [see \Eq{A LL 8}]. 

We seek the solution of \Eq{pt 1} in the form of an expansion inspired by \Eq{pt 4}, 
\beq
\varrho(t,\zeta) = \varrho_0(t) + \frac{\pi\zeta}{6}\varrho_1(t) + \ldots,
\label{pt 6}
\eeq
where the functions $\varrho_0(t)$ and $\varrho_1(t)$ are independent of $\zeta$. Since $\varrho(0,\zeta) = 1$ for any $\zeta$, these functions satisfy 
\beq
\varrho_0(0) = 1,
\quad
\varrho_1(0) = 0.
\label{pt 7}
\eeq
On the other hand, we expect the functions $\varrho(t,\zeta)$ and $\widetilde\varrho(t,\zeta)$ to match at $1\ll |t|\ll1/\zeta$ in every order in $\zeta$. This leads to the relations
\beq
\lim_{|t|\,\to\,\infty}\frac{\varrho_0(t)}{\widetilde\varrho_0(t)} = 1,
\quad
\lim_{|t|\,\to\,\infty}\frac{\varrho_1(t)}{\widetilde\varrho_1(t)} = 1,
\label{pt 8}
\eeq
so that $\varrho_1^\prime(t) = d\varrho_1(t)/dt$ satisfies
\beq
\lim_{|t|\,\to\,\infty}\frac{\varrho_1^\prime(t)}{\widetilde\varrho_0(t)} = 1,
\label{pt 9}
\eeq
where we used $\widetilde\varrho_1^{\,\prime}(t) = \widetilde\varrho_0(t)$.

Formally, the right-hand side of \Eq{pt 6} represents an asymptotic expansion of $\varrho(t,\zeta)$ in small $\zeta$ evaluated at fixed~$t$. The expansion is applicable as long as the second term is small compared with the first one. With Eqs.~\eq{pt 7} and \eq{pt 8} taken into account, this yields the condition \mbox{$\zeta \ll \min\{1,|t|^{-1}\}$} on~$\zeta$, which translates to $|t|\ll1/\zeta$ for~$t$. Taking advantage of this inequality, we introduce the intermediate scale $\tau$ satisfying
\beq
\max\bigl\{1,|t|\bigr\}\ll\tau\ll 1/\zeta,
\label{pt 10}
\eeq 
break the integral in $\int_{-2k_0}^0\!\!dt'\Theta^\prime(t \!-\! t')\varrho(t')$ in two, \mbox{$\int^0_{-2k_0}\!dt'[\ldots] = \int^0_{-\tau}\!dt'[\ldots] + \int^{-\tau}_{-2k_0}\!dt'[\ldots]$}, and replace $\Theta$ and $\rho$ in the second integral here with $\widetilde\Theta$ and $\widetilde\varrho$. (The relative error of such an approximation is of order $1/\tau\ll 1$.) With the help of Eqs.~\eq{pt 1}-\eq{pt 3}, we obtain the equation
\beq
\mathcal L_{\tau}[\varrho] = \widetilde{\mathcal L}_{\tau}[\widetilde\varrho\,]. 
\label{pt 11}
\eeq

The functions $\varrho_0(t)$ and $\varrho_1(t)$ in \Eq{pt 6} can now be deduced by replacing $\widetilde\varrho$ and $\varrho$ in \Eq{pt 11} with the expansions \eq{pt 4} and \eq{pt 6}, respectively, and considering the behavior of the resulting equation at small $\zeta$ and large~$\tau$. For the equation to hold at large, but still finite $\tau$ and arbitrarily small $\zeta\ll 1/\tau$ [see \Eq{pt 10}], it must be satisfied in every order in $\zeta$ separately. This observation leads to the equations
\begin{subequations}
\label{pt 12}
\beqa
\mathcal L_{\tau}[\varrho_0] &=& \widetilde{\mathcal L}_{\tau}[\widetilde\varrho_0],
\label{pt 12a} \\
\mathcal L_{\tau}[\varrho_1] &=& \widetilde{\mathcal L}_{\tau}[\widetilde\varrho_1].
\label{pt 12b}
\eeqa
\end{subequations}
Differentiating these equations with respect to $\tau$, we recover the relations \eq{pt 8} and \eq{pt 9}.

In the limit $\tau\to\infty$, \Eq{pt 12a} yields
\beq
\mathcal L_\infty[\varrho_0] = 0,
\label{pt 13}
\eeq
where we took into account \Eq{pt 5}. Equation~\eq{pt 13} coincides with \Eq{BA 24}. The solution of this equation satisfying the condition $\varrho_0(0) = 1$ [see \Eq{pt 7}] is unique and behaves at large $|t|$ as prescribed by the first equation in \eq{pt 8}, see \Sec{evaluation} and Appendix~\ref{Appendix_Wiener-Hopf}. 

Because the integrals in \Eq{pt 12b} diverge at \mbox{$\tau\to\infty$}, we first differentiate both sides of this equation with respect to $t$ and integrate by parts using \mbox{$\varrho_1(0) = \widetilde\varrho_1(-0) = 0$}. Taking now the limit $\tau\to\infty$ and using \Eq{pt 5}, we obtain the equation
\beq
\mathcal L_\infty[\varrho_1^\prime] = 0.
\label{pt 14}
\eeq
Its solution subject to the condition \eq{pt 9} reads \mbox{$\varrho_1^\prime(t) = \varrho_0(t)$}. Combining this result with the second equation in~\eq{pt 7}, we finally arrive at 
\beq
\varrho_1(t) = \int_0^t dt'\varrho_0(t').
\label{pt 15}
\eeq
 
By construction, \Eq{pt 6} with $\varrho_1(t)$ given by \Eq{pt 15} has the form of \Eq{A LL 7}. We have verified that Eqs.~\eq{pt 6} and \eq{pt 15} with $\varrho_0(t)$ approximated by its asymptote~\eq{BA 31b} are in agreement with the expansion derived by a different method in Ref.~[\onlinecite{Popov_LL}]. 

\section{Solution of Eq.~(\ref{BA 24})}
\label{Appendix_Wiener-Hopf}

In this Appendix we employ the Wiener-Hopf technique (see, e.g., Ref.~[\onlinecite{MorseFeshbach}]) to construct the solution of \Eq{BA 24}. 

Substituting $\varrho_0(t)$ in the form
\beq
\varrho_0(t) = \varrho_+(t) + \varrho_-(t), 
\quad
\varrho_\pm(t) = \theta(\pm \,t)\,\varrho_0(t)
\label{WH 1}
\eeq
into \Eq{BA 24}, we obtain the equation
\beq
\varrho_+(t) + \int_{-\infty}^{\infty}\!\!dt'\mathcal G(t-t')\varrho_-(t') = 0
\label{WH 2}
\eeq
with
\beq
\mathcal G(t) = \delta(t) + \frac{\Theta'(t)}{2\pi\,}
=  \delta(t) - \frac{1\,}{\pi(1 + t^2)}.
\label{WH 3}
\eeq
Upon the Fourier transform, \Eq{WH 2} assumes the form
\beq
\varrho_+(\omega) + \mathcal G(\omega)\varrho_-(\omega) = 0.
\label{WH 4}
\eeq
The functions
\beq
\varrho_\pm(\omega) = \int_{-\infty}^\infty\!dt\,e^{i\omega t} \varrho_\pm(t)
\label{WH 5}
\eeq
in \Eq{WH 4} are analytic at $\pm\,\text{Im}\,\omega \geq 0$. Their behavior at large $|\omega|$ is obtained by substituting $\varrho_0(t)$ in the form of the Taylor series \mbox{$\varrho_0(t) = 1 + \varrho_0^{\,\prime}(0)t + \ldots$} [recall that $\varrho_0(0) = 1$] into Eqs.~\eq{WH 1} and \eq{WH 5}, which yields
\beq
\varrho_\pm(\omega)\bigr|_{|\omega|\,\gg\,1} = \pm\,\frac{i}{\omega}\left[1 + \varrho_0^\prime(0)\frac{i}{\omega} + \ldots\right].
\label{WH 6}
\eeq

The kernel $\mathcal G(\omega)$ in \Eq{WH 4} is given by
\beq
\mathcal G(\omega) = \int_{-\infty}^\infty\!dt\,e^{i\omega t}\,\mathcal G(t)
= 1 - e^{-|\omega|}.
\label{WH 7}
\eeq
It can be factorized as
\beq
\mathcal G(\omega) = -\,\frac{F_+(\omega)}{F_-(\omega)}
\label{WH 8}
\eeq
with
\beq
F_+(\omega) = \widetilde\varrho_+(\omega) f_+(\omega),
\quad
F_-(\omega) = \frac{\widetilde\varrho_-(\omega)}{f_-(\omega)}.
\label{WH 9}
\eeq
Here
\beq
\widetilde\varrho_+(\omega) = \frac{e^{i\pi/4}~~}{\,(\omega + i0)^{1/2}}\,,
\quad
\widetilde\varrho_-(\omega) = \frac{e^{-3i\pi/4}~~}{\,(\omega - i0)^{3/2}}
~~
\label{WH 10}
\eeq
are Fourier transforms of $\widetilde\varrho_\pm(t) = \theta(\pm\, t)\widetilde\varrho_0(t)$, where
$\widetilde\varrho_0(t)$ is the solution \eq{A LL 8} of the approximate Lieb equation~\eq{pt 5}. The functions $\widetilde\varrho_\pm(\omega)$ are analytic in the complex plane with the branch cuts running from $\mp\,i0$ to $\mp \,i\infty$.
The functions $f_\pm(\omega)$ in \Eq{WH 9} are given by
\beq
f_\pm(\omega) 
= \frac{\exp\!\left\{\,\mp\,i\frac{\omega}{2\pi}\!\left[\,\ln\!\left(\frac{\omega\,\pm \,i0}{2\pi}\right)-1 \,\mp\, i\frac{\pi}{2}\right]\right\}}
{\Gamma\!\left(1\mp \,i\frac{\omega}{2\pi}\right)},
\label{WH 11}
\eeq
with the same branch cut structure. These functions approach~$1$ at $\omega\to 0$, whereas at large $|\omega|$ application of the Stirling formula gives the asymptotes
\beq
f_\pm(\omega)\bigr|_{|\omega|\,\gg\,1} 
= \frac{e^{\pm \,i\pi/4}}{\sqrt{\omega\,}} 
\left(
1\mp\frac{i\pi}{6_{}\omega} + \ldots
\right)\!.
\label{WH 12}
\eeq

Because $\widetilde\varrho_\pm(\omega)$ and $f_\pm(\omega)$ have no singularities or zeros at \mbox{$\pm\,\text{Im\,}\omega \geq 0$}, the functions $F_\pm(\omega)$ given by \Eq{WH 9} are analytic in these regions. Standard arguments\cite{MorseFeshbach} then show that the function
\beq
\mathcal F(\omega) = \frac{\varrho_+(\omega)}{F_+(\omega)} 
=  \frac{\varrho_-(\omega)}{F_-(\omega)}
\label{WH 13}
\eeq
is analytic in the entire complex plane. Using Eqs.~\eq{WH 6}, \eq{WH 9}, \eq{WH 10}, and \eq{WH 12}, we find $\mathcal F(\omega)\bigr|_{|\omega|\to\infty} = 1$. Therefore, $\mathcal F = 1$ at all $\omega$, and \Eq{WH 13} yields
\beq
\varrho_\pm(\omega) = F_\pm(\omega).
\label{WH 14}
\eeq

At small $\omega$, \Eq{WH 14} reduces to \mbox{$\varrho_\pm(\omega)\bigr|_{|\omega|\,\ll\, 1} = \widetilde\varrho_\pm(\omega)$}, hence at $|t|\gg 1$ the inverse Fourier transforms
\beq
\varrho_\pm(t) = \int_{-\infty}^\infty\frac{d\omega}{2\pi}\,e^{- i\omega t}\varrho_\pm(\omega)
\label{WH 15}
\eeq
reproduce the classical asymptotes \Eq{BA 19}. At large~$\omega$, \Eq{WH 14} yields the expansion \eq{WH 6} with
\beq
\varrho_0^{\,\prime}(0) = -\,\pi/6,
\label{WH 16}
\eeq
resulting in the Taylor series \eq{BA 35}. At intermediate $\omega$ and $t$ no further simplifications are possible. Instead, we deform the integration paths in \Eq{WH 15} to run along the respective branch cuts and change the integration variables to $z = \pm\,i\omega/2\pi$. Taking into account \Eq{WH 1}, we arrive at Eqs.~\eq{BA 30}. 

\section{Derivation of Eq.~(\ref{BA 27})}
\label{LL vs QT}

In this Appendix, we derive the relation \eq{BA 27} between the functions $\varrho_0(t)$ for the Lieb-Liniger model and for the quantum Toda model.

Substituting 
\beq
\varrho_0(t) = \bar\varrho_0(-t)
\label{LL vs QT 1}
\eeq
\\
into \Eq{BA 26} and using \mbox{$\Theta^{\prime\prime}(t) = -\,\Theta^{\prime\prime}(-t)$} [see \Eq{BA 25}], we obtain the equation
\beq
\bar\varrho_0^{\,\prime}(t) + \frac{1}{2\pi} \int_0^\infty\!dt'\Theta^{\prime\prime}(t-t')\bar\varrho_0(t') = 0.
\label{LL vs QT 2}
\eeq
Writing $\bar\varrho_0(t)$ as
\beq
\bar\varrho_0(t) = \bar\varrho_+(t) + \bar\varrho_-(t),
\quad
\bar\varrho_\pm(t) = \theta(\pm \,t)\bar\varrho_0(t)
\label{LL vs QT 3}
\eeq
[cf.~\Eq{WH 1}], and using the continuity of $\bar\varrho_0(t)$ at $t = 0$, we find
\beq
\bar\varrho_-^{\,\prime}(t) + \int_{-\infty}^\infty\!dt'\bar{\mathcal G}^{\,\prime}(t-t')\bar\varrho_+(t') = 0.
\label{LL vs QT 4}
\eeq
Here
\beq
\overline{\mathcal G}(t)
= \delta(t) - \,\frac{1}{\pi}\,\text{Re}\,\psi(1 + it),
\label{LL vs QT 5}
\eeq
where $\psi(z)$ is the digamma function. Fourier transform of \Eq{LL vs QT 5} reads
\beq
\overline{\mathcal G}(\omega) = \int_{-\infty}^\infty\!dt\,e^{i\omega t} \,\overline{\mathcal G}(t) 
= \frac{1}{1 - e^{-|\omega|}} 
= \frac{1}{\mathcal G(\omega)}\,,
\label{LL vs QT 6}
\eeq
with $\mathcal G(\omega)$ given by \Eq{WH 7}.
With the help of \Eq{LL vs QT 6}, Fourier transform of \Eq{LL vs QT 4} can then be written as
\beq
\bar\varrho_+(\omega) + \mathcal G(\omega)\bar\varrho_-(\omega) = 0.
\label{LL vs QT 8}
\eeq
Carrying out the inverse Fourier transform, we obtain
\beq
\bar\varrho_+(t) + \int_{-\infty}^\infty\!\!dt'\mathcal G(t-t')\bar\varrho_-(t')  = 0,
\label{LL vs QT 9}
\eeq
which is the same as \Eq{WH 2}. The corresponding equation for $\bar\varrho_0(t)$ then coincides with \Eq{BA 24} for the Lieb-Liniger model and is subject to the same constraint $\bar\varrho(0) = 1$ at $t = 0$. Therefore, $\bar\varrho_0(t)$ for the quantum Toda model is identical to $\varrho_0(t)$ for the Lieb-Liniger model, which leads to \Eq{BA 27}. 



\end{document}